\batchmode
\makeatletter
\def\input@path{{D:/Australia/thz/meeting/codetens/}}
\makeatother
\documentclass[twocolumn,journal]{IEEEtran}
\usepackage[T1]{fontenc}
\usepackage[latin9]{inputenc}
\usepackage{array}
\usepackage{float}
\usepackage{textcomp}
\usepackage{mathrsfs}
\usepackage{amsmath}
\usepackage{amsthm}
\usepackage{amssymb}
\usepackage{stackrel}
\usepackage{graphicx}
\usepackage{cite}
\usepackage[pdftex,unicode=true,
 bookmarks=true,bookmarksnumbered=true,bookmarksopen=true,bookmarksopenlevel=1,
 breaklinks=false,pdfborder={0 0 0},pdfborderstyle={},backref=false,colorlinks=false]
 {hyperref}
\hypersetup{pdftitle={Your Title},
 pdfauthor={Your Name},
 pdfpagelayout=OneColumn, pdfnewwindow=true, pdfstartview=XYZ, plainpages=false}

\makeatletter

\newcommand{\lyxmathsym}[1]{\ifmmode\begingroup\def\b@ld{bold}
  \text{\ifx\math@version\b@ld\bfseries\fi#1}\endgroup\else#1\fi}

\providecommand{\tabularnewline}{\\}
\floatstyle{ruled}
\newfloat{algorithm}{tbp}{loa}
\providecommand{\algorithmname}{Algorithm}
\floatname{algorithm}{\protect\algorithmname}

\let\oldforeign@language\foreign@language
\DeclareRobustCommand{\foreign@language}[1]{%
  \lowercase{\oldforeign@language{#1}}}
\theoremstyle{plain}
\newtheorem{thm}{\protect\theoremname}
\theoremstyle{plain}
\newtheorem{lem}[thm]{\protect\lemmaname}

\usepackage[caption=false,font=footnotesize]{subfig}

\@ifundefined{showcaptionsetup}{}{%
 \PassOptionsToPackage{caption=false}{subfig}}
\usepackage{subfig}
\makeatother

\providecommand{\lemmaname}{Lemma}
\providecommand{\theoremname}{Theorem}

\begin{document}
\title{Quantum Sensing Based Joint 3D Beam Training for UAV-mounted STAR-RIS
Aided TeraHertz Multi-user Massive MIMO Systems }
\author{Xufang~Wang,~\IEEEmembership{Member,~IEEE,} Zihuai~Lin,~\emph{Senior
Member,~IEEE},~Feng~Lin,~\IEEEmembership{Member,~IEEE,}~Pei~Xiao,~\IEEEmembership{Senior Member,~IEEE}\thanks{Xufang Wang is with the Key Laboratory of Optoelectronic Science and
Technology for Medicine of Ministry of Education, Fujian Normal University,
Fuzhou, China (e-mail: fzwxf@fjnu.edu.cn). }\thanks{Zihuai Lin is with the School of Electrical and Information Engineering,
University of Sydney, Sydney, NSW 2006, Australia (e-mail: zihuai.lin@sydney.edu.au).}\thanks{Feng~Lin is with Kongtronics Institute of Science and Technology
(XiaMen) Co., Ltd., Xia Men, China (e-mail: ffglinfeng@163.com).}\thanks{Pei Xiao is with the Institute for Communication Systems (ICS), home
of 5GIC \& 6GIC, University of Surrey, UK (e-mail: p.xiao@surrey.ac.uk
)}}
% \markboth{IEEE TRANSACTIONS ON WIRELESS COMMUNICATIONS }{Xufang Wang \MakeLowercase{\emph{et al.}}: Quantum Sensing Based
% joint 3D Beam Training for UAV-mounted STAR-RIS Aided TeraHertz Multi-user
% Massive MIMO systems}
\maketitle
\begin{abstract}
\emph{Terahertz} (THz) systems are capable of supporting ultra-high
data rates thanks to large bandwidth, and the potential to harness
high-gain beamforming to combat high pathloss. In this paper, a novel
quantum sensing\emph{\textemdash Ghost Imaging} (GI) based beam training
is proposed for \emph{Simultaneously Transmitting and Reflecting}
\emph{Reconfigurable Intelligent Surface} (STAR RIS) aided THz multi-user
massive MIMO systems. We first conduct GI by surrounding 5G downlink
signals to obtain the 3D images of the environment including users
and obstacles. Based on the information, we calculate the optimal
position of the UAV-mounted STAR by the proposed algorithm. Thus the
position-based beam training can be performed. To enhance the beamforming
gain, we further combine with channel estimation and propose a semi-passive
structure of the STAR and ambiguity elimination scheme for separated
channel estimation. Thus the ambiguity in cascaded channel estimation,
which may affect optimal passive beamforming, is avoided. The optimal
active and passive beamforming are then carried out and data transmission
is initiated. The proposed BS sub-array and sub-STAR spatial multiplexing
architecture, optimal active and passive beamforming, digital precoding,
and optimal position of the UAV-mounted STAR are investigated jointly
to maximize the average achievable sum-rate of the users. Moreover,
the \emph{cloud radio access networks} (CRAN) structured 5G downlink
signal is proposed for GI with enhanced resolution. The simulation
results show that the proposed scheme achieves beam training and separated
channel estimation efficiently, and increases the spectral efficiency
dramatically compared to the case when the STAR operates with random
phase. Furthermore, the proposed scheme improves the spectral efficiency
on average by 14.26\%, 18.35\% and 60.60\% for three different configurations,
respectively, compared to that at the 10\% deviation position even
with perfect CSI.
\end{abstract}

\begin{IEEEkeywords}
TeraHertz, quantum sensing, Ghost Imaging, hybrid 3D beam training,
channel estimation, Reconfigurable Intelligent Surface
\end{IEEEkeywords}

\IEEEpeerreviewmaketitle{}

\section{Introduction}

\IEEEPARstart{T}{}erahertz (THz) communications have been regarded
as a promising candidate to support the explosive growth of mobile
devices and seamless multimedia applications for the future 6G wireless
networks. Although THz communications have the advantage of increasing
the bandwidth by orders of magnitude, they inherently suffer from
limited coverage due to the path loss incurred at high frequencies,
the absorption of molecules in the atmospheric medium and the higher
probability of \emph{line-of-sight} (LOS) blockage \cite{ma_joint_2020}.
On the other hand, they tend to harness high-gain or extremely narrow
beam, which requires high accuracy of beam pointing to the receiver.
In \cite{lin_energy-efficient_2016}, a hybrid beamforming system
associated with an array-of-subarray structure is proposed for THz
communications. To further alleviate the short-range bottleneck of
THz network, a feasible cost-effective approach is to integrate it
with a \emph{Reconfigurable Intelligent Surface} (RIS), whose phase
shift can be controlled by a low-complexity programmable PIN diode
\cite{pan_intelligent_2020}. RIS can provide virtual LoS paths and
reduce the probability of blockage to improve the THz propagation
conditions \cite{wu_intelligent_2019,xu2021THzUAV,Liu2022THzRIS,du2022ITS_RIS,chu2022IoT_RIS}, but the transmitter, RIS and
receiver need to be co-designed for beam training and tracking with
interference suppression taken into account.

Although the optimization of most RIS-aided MISO systems can be directly
transformed into quadratic programming under quadratic constraints,
there is a paucity of literature on the design of RIS assisted MIMO
systems \cite{zhang_capacity_2020,ye_joint_2020}. Additionally, most
of the open literature assumes that the \emph{channel state information}
(CSI) of the RIS is perfectly available \cite{wu_intelligent_2019}\cite{hu_performance_2021}\cite{guo_weighted_2020}.
However, since all elements of the RIS are passive, it cannot send,
receive or process any pilot signal for channel estimation. Moreover,
RISs usually contain hundreds of elements, and the dimension of the
estimated channel is thus much larger than that of traditional systems,
resulting in an excessive pilot overhead. Therefore, traditional solutions
cannot be directly applied, and channel estimation is a key challenge
\cite{wei_channel_2021}. In this regard, sophisticated schemes have
been proposed for the channel estimation of RIS-aided MISO systems
\cite{ma_joint_2020} \cite{elbir_deep_2020} \cite{shaham_fast_2019}
\cite{liu_deep_2020}, where the receiver is equipped with a single
antenna. However, these schemes cannot be readily combined with the
above-mentioned channel estimation schemes in massive MIMO systems \cite{Chen2021MIMO,MIMO_capacity,Chen2021MIMO,network_capacity}.
Currently, the cascaded BS-RIS-user channel is generally chosen for
channel estimation, e.g., the tensor based channel estimation \cite{de_araujo_channel_2021}
based on AWGN channels, in which, inherently unavoidable ambiguities
of channels exist. These ambiguities undermine the application of
RISs in many wireless services such as localization or mobility tracking.
Moreover, traditional beam training methods are unsuitable for the
extremely narrow pencil beams of THz waves. A cooperative beam training
scheme is developed in \cite{ning_terahertz_2021} to facilitate the
estimation of the concatenated twin-hop BS-RIS-user channel. However,
they assumed having no obstacles between the BS and users, which may
not always be the case. 

On the other hand, the integration of sensing functionality has become
a key feature of the 6G \emph{Radio Access Network} (RAN) and \emph{Integrated
Sensing and Communication} (ISAC) \cite{liu_integrated_2022} have
attracted substantial research attention. In this context, a number
of ISAC schemes have been proposed. For example, the co-existence
and joint transmission for a MIMO \emph{Radar-Communication }(RadCom)
system is proposed in \cite{liu_mu-mimo_2018} subject to specific
\emph{Signal to Interference and Noise Ratio} (SINR) constraints.
A Bayesian prediction based low-overhead joint radar-communication
approach is proposed in \cite{yuan_bayesian_2021} for Vehicular Networks.
To tackle the 3D beam training problem for STAR aided THz multi-user
massive MIMO networks, we will not only propose a novel ISAC scheme
based on quantum sensing, but also propose an algorithm to obtain
the optimal position of the UAV-mounted STAR, as UAVs can be maneuvered
to the desired location, which provides new \emph{degrees of freedom}
(DOF) to optimize performance.

A quantum sensing technique named \emph{Ghost Imaging} (GI), which
is originated from quantum and optics, has recently attracted much
attention \cite{wang_microwave_2016,GI1,GI4,GI_ICASSP,GI_ECC,GI_Micro} because of its unique features,
e.g., nonlocal reconstruction, non-scanning, super-resolution, etc.
The original GI experiments using the entanglement properties of photon
pairs, subsequent studies have found that entangling light and hot
(turns) light can realize correlation imaging \cite{dongze_li_radar_2014}.
The beam splitter is divided into two optical paths: reference light
and object arm light. The reference light is recorded by a \emph{charge-coupled
device} (CCD), and object arm light is received by barrel detector
without spatial resolution. According to the intensity fluctuation
correlation theory, the object arm light and the reference light are
connected to restore the object image. 

As light is basically an \emph{electromagnetic} (EM) wave, GI has
been recently modified and integrated into microwave imaging. As a
reference signal, the microwave radiation field can be obtained either
by recording or measuring the signal or by deducing and calculating
the signal model. Therefore, microwave association imaging is easier
to be implemented by computational imaging, and can be combined with
advanced signal processing algorithms in imaging processing to improve
imaging performance \cite{dongze_li_radar_2014}. A LTE-based microwave
GI system is proposed in \cite{zhang_microwave_2018}, which can reconstruct
objects effectively. Without deliberate deployment of transmitters
and receivers, the complexity and cost of the microwave GI system
can be greatly reduced. However, the GI resolution in \cite{zhang_microwave_2018}
is in hundreds of meters. To further improve the GI resolution, we
propose a novel quantum sensing 5G GI system for environment including
obstacle and user detection, which can be further utilized for UAV-mounted
STAR-aided THz beamforming. 

For RIS-aided networks, the received signal suffers from ``double
fading'', i.e., the propagation loss both from the BS-RIS link and
RIS-user link, which largely relies on RIS positions. Especially for
THz signal, even some additional path loss due to inaccurate RIS positions
can greatly affect the received signal strength, thereby weakening
the RIS gain. Although RIS coordinates was discussed in \cite{wu_intelligent_2019},
RIS deployment concerning important influencing factors such as communication
environment including obstacles (e.g., buildings, which usually have
certain dimensions and should not be ignored) have never been considered
to date in literature. For example, a RIS can not be deployed inside
an obstacle. At least the edges of the obstacles should be detected
and used as references for the deployment of the RISs to optimize
performance due to the severe path loss of THz signals. To fill the
gap, we formulate the problem as how to jointly design the THz BS
and STAR architectures, deployment of beamforming (both active and
passive) to achieve the optimal performance of multi-user THz massive
MIMO systems in diverse communication environments with obstacles.
Furthermore, we also adopt a new category of RISs called \emph{Simultaneously
Transmitting and Reflecting} RISs (STAR-RISs) \cite{niu_simultaneous_2021},
which can serve users distributed at both sides of RISs. For the sake
of brevity, we simplify STAR-RISs as STARs. 

In this paper, we propose a novel GI based joint 3D beam training
scheme for the UAV-mounted STAR aided THz multi-user MIMO Systems.
We first conduct GI by 5G downlink signals to obtain the 3D images
of the environment including users and obstacles such as building,
tower, etc. Then we perform K-means to obtain the user clustering
and centroids of the user clusters. Based on these information, we
calculate the optimal position of the UAV-mounted STAR by the proposed
algorithm. Subsequently, the position-based beam training can be carried
out. To enhance the beamforming gain, we further combine with channel
estimation and propose a semi-passive structure of the STAR and ambiguity
elimination scheme for separated channel estimation. Thus the ambiguity
in cascaded channel estimation, which may affect optimal passive beamforming,
is avoided. The optimal active and passive beamforming are then carried
out and data transmission is initiated. In addition, an array-of-subarray
based THz BS architecture and the corresponding subSTAR structure
are carefully designed to cooperate for spatial multiplexing. The
BS, STAR and receiver antenna arrays of the users are all uniform
planar arrays (UPAs). The joint 3D beamforming architecture can be
extended to multiple-STAR scenario. Moreover, the \emph{cloud radio
access networks} (CRAN) structured 5G downlink signal is proposed
for GI with enhanced resolution. The simulation results show that
the proposed scheme achieves beam training and separated channel estimation
efficiently, and increases the spectral efficiency dramatically compared
to the scenario where the STAR operates with random phase.

Against the above backdrop, our main contributions are summarized
below.

1) We propose a novel GI based joint hybrid 3D beamforming architecture
for UAV-mounted STAR-aided THz multi-user massive MIMO systems. The
CRAN structured 5G downlink signal is proposed for GI with enhanced
resolution. The proposed ISAC sytem is inherently integrated, resulting
in low operational cost.

2) We also conceive a GI based beam training scheme combined with
a separated channel estimator as well as a semi-passive STAR architecture,
which eliminates the ambiguity in cascaded channel estimation that
may affect optimal passive beamforming. 

3) To achieve high system performance, we propose a 3D optimal position
finding algorithm for UAV-mounted STARs in the communication environment
where dimensions of obstacles are taken into account for the first
time. 

4) The proposed system is scalable since its performance is not affected
by the number of users as there is almost no interference among each
pair of BS subarray-subSTAR-user. Moreover, as the operation of the
system can be carried out in parallel, the algorithm complexity will
not be affected by the number of users. In addition, the BS subarrays
and subSTARs of the proposed architecture can be flexibly or dynamically
configured to meet different needs of users.

5) The proposed BS sub-array and sub-STAR spatial multiplexing architecture,
optimal active and passive beamforming, digital precoding, and optimal
position of the UAV-mounted STAR are investigated jointly to maximize
the average achievable sum-rate of the users. 

6) Our proposed 3D beam training scheme is not restricted by GI and
can be used in combination with other sensing approaches. The scheme
has practical potential for emerging STAR aided THz applications such
as integrated networks of terrestrial links, UAVs, and satellite communication
systems. 

The remainder of the paper is organized as follows. In Section II,
we describe the system model and formulate the optimization problems.
In Section III, we explore the conditions of achieving high-integrity
spatial multiplexing for the proposed STAR-aided THz architecture.
In Section IV, the proposed quantum sensing based joint 3D beam training
scheme is presented for the UAV-mounted STAR aided THz multi-user
MIMO Systems. The simulation results and discussion are given in Section
V. Finally, we conclude in Section VI. 

\emph{Notation: }Boldface lower case and upper case letters are used
for column vectors and matrices, respectively. The superscripts $(\cdot)^{*}$
, $(\cdot)^{T}$, $(\cdot)^{H}$, and $(\cdot)^{-1}$ stand for the
conjugate, transpose, conjugate-transpose, and matrix inverse, respectively.
The Euclidean norm, absolute value, Hadamard product are denoted by
$||\cdot||$, $|\cdot|$ and $\odot$ respectively. In addition, $\mathbb{E}\left\{ \cdot\right\} $
is the expectation operator. For a matrix $\mathbf{A}$, $[\mathbf{A}]_{mn}$
denotes its entry in the \emph{m}-th row and \emph{n}-th column, while
for a vector $\mathbf{a}$, $[\mathbf{a}]_{m}$ denotes the \emph{m}-th
entry of it. Furthermore, $j$ in $e^{j\theta}$ denotes the imaginary
unit, while $\mathbf{I}$ is the identity matrix. 

\section{System Model and Problem Formulation}

In this section, we introduce the system and channel models of 3D
hybrid beamforming designed for UAV mounted STAR-assisted THz MIMO
systems, including the direct BS-to-user path and the BS-STAR-user
path. Then we formulate the optimization problem to be solved.

\subsection{System Model}

\begin{figure*}[tbh]
\centering{}\includegraphics[scale=0.45]{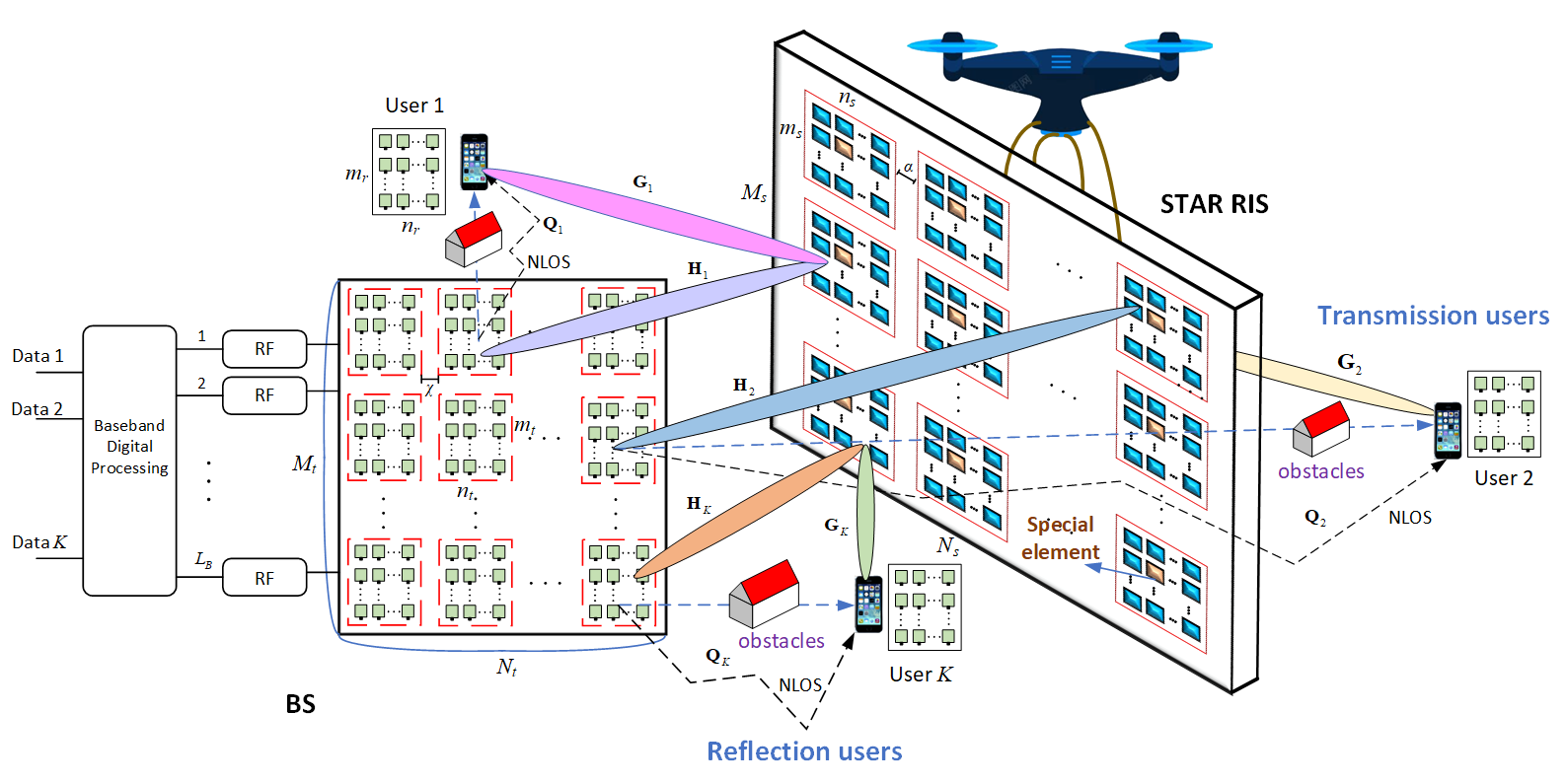}\caption{\label{fig:System-model-of}System model of the novel spatial multiplexing
architecture for UAV-mounted STAR-aided THz multi-user massive MIMO
systems.}
\end{figure*}

The communication system under study is depicted in Fig. \ref{fig:Communication-Scenario}.
It is worth mentioning that in the system model for the BS-STAR-user
path, a novel joint hybrid 3D beamforming BS-subarray and the corresponding
subSTAR architecture are proposed for UAV-mounted STAR-aided THz multi-user
massive  MIMO systems, the realization of which will be detailed in
the following sections. The system model we adopted is shown in Fig.
\ref{fig:System-model-of}. The THz transceivers have array-of-subarrays
of graphene-based plasmonic nano-antennas \cite{akyildiz_realizing_2016}.
The BS transmitter (TX) having $L_{B}=M_{t}\times N_{t}$ subarrays
supports \emph{$K$} users either with or without STARs. The \emph{i}th
subarray of the BS is a UPA having $m_{t,i}\times n_{t,i}$ antenna
units. For simplicity and without loss of generality, we let $m_{t,i}=m_{t}$,
$n_{t,i}=n_{t},$ $i=1,...,L_{B}$. Note that $L_{B}$ is also the
number of RF chains, since each BS subarray is controlled by a dedicated
RF chain. Due to limited processing power, there is only a single
subarray baseband and RF chain consisting of $m_{r,k}\times n_{r,k}$
tightly-packed elements at the \emph{k}th user. For simplicity and
without loss of generality, we let $m_{r,k}=m_{r}$, $n_{r,k}=n_{r},$
$k=1,...,K$. The number $L_{B}$ of antenna subarrays is assumed
to be higher than \emph{$K$} for attaining high gains. UPAs are promising
for THz communications both in BS and user terminals, since they can
accommodate more antenna elements by a two-dimensional subarray for
3D beamforming.
\begin{center}
\begin{figure}[tbh]
\centering{}\includegraphics[scale=0.5]{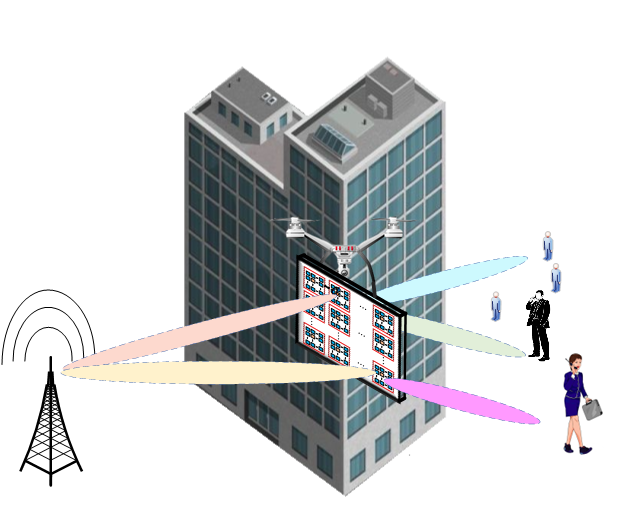}\caption{\label{fig:Communication-Scenario}UAV-mounted STAR aided THz multi-user
massive MIMO systems.}
\end{figure}
\par\end{center}

Blockage is commonplace in THz communications. Hence we assume that
the line-of-sight (LOS) link from the BS to each user is indeed blocked,
and a STAR is applied for improving the link spanning from the BS
to the user by reflecting and refracting the signal. The STAR consists
of a sub-wavelength UPA having $\varUpsilon$ passive reflecting and
transmitting elements under the control of a STAR controller. The
THz channel is highly frequency-selective, but there are several low-attenuation
windows separated by high-attenuation spectral nulls owing to molecular
absorption \cite{moldovan_and_2014}. Therefore, we can adaptively
divide the total THz signal bandwidth into numerous sub-bands, say
$C$ sub-bands. The \emph{h}th sub-band is centered around frequency
$f_{h},h=1,2,\ldots,C$ and it has a width of $\varPi_{f_{h}}$. If
$\varPi_{h}$ is small enough, the channel can be regarded as frequency-non-selective
and the noise power spectral density appears to be locally flat. Thus,
for the \emph{h}th subband signal, we will discuss the direct path
spanning from the BS to the user and that from the BS to the user
via STAR, i.e., the BS-STAR-user path.

(1) \textbf{\emph{The direct BS-user path}}

The received signal of user $k$ can be expressed as

\begin{equation}
\bar{y}_{k}=\mathbf{v}_{k}^{H}\mathbf{Q}_{k}\mathbf{W}\mathbf{Fs}+\mathbf{v}_{k}^{H}\mathbf{n}_{k},~k=1,...,~K,\label{eq:direct}
\end{equation}
where $\mathbf{s}=[s_{1},s_{2},...,s_{K}]^{T}$ is a $K\times1$ vector
containing the transmitted symbols of $K$ users, so that $\mathbb{E}\left\{ \mathbf{ss}^{*}\right\} =\frac{P}{K}\mathbf{I}_{K}$,
where $P$ denotes the total initial transmit power and the same power
is assigned to each user. In (\ref{eq:direct}), $\mathbf{Q}_{k}$
is the $m_{r}n_{r}\times L_{B}m_{t}n_{t}$ THz channel matrix between
the BS and user $k$. The LOS components of the direct BS-user links
are blocked by obstacles, we thus assume that the direct link channel
$\mathbf{\mathbf{Q}_{\mathit{k}}}$ contains only \emph{non-line-of-sight}
(NLOS) components; $\mathbf{W}$ is the $L_{B}m_{t}n_{t}\times L_{B}$
analog transmit beamforming matrix representing the equal power splitters
and phase shifters. For the array-of-subarray structure of (\ref{eq:direct}),
$\mathbf{W}$ is block-diagonal structure and can be expressed as

\begin{equation}
\mathbf{W}=\left[\begin{array}{cccc}
\mathbf{w}_{1} & \mathbf{0} & \cdots & \mathbf{0}\\
\mathbf{0} & \mathbf{w}_{2} & \cdots & \mathbf{0}\\
\vdots & \vdots & \vdots & \vdots\\
\mathbf{0} & \mathbf{0} & \cdots & \mathbf{w}_{L_{B}}
\end{array}\right],
\end{equation}
where $\mathbf{w}_{l}$ is an $m_{t}n_{t}\times1$ vector; let $\mathbf{\mathbf{w}_{\mathit{l}}\left(\mathit{i}\right)}$
denote the $i_{th}$ element of $\mathbf{w}_{\mathit{l}}$, $|\mathbf{w}_{\mathit{l}}\left(\mathit{i}\right)|=\frac{1}{\sqrt{m_{t}n_{t}}},$
$l=1,...,L_{B}$, $i=1,...,m_{t}n_{t}$. Still referring to (1), $\mathbf{F}$
is the $L_{B}\times K$ baseband digital beamforming matrix used for
interference mitigation, and can be expressed as
\begin{equation}
\mathbf{F}=\left[\mathbf{f}_{1},\mathbf{\mathbf{f}}_{2},\ldots,\mathbf{\mathbf{f}}_{K}\right],
\end{equation}
with 
\begin{equation}
||\mathbf{W}\mathbf{f}_{k}||^{2}=1;\label{eq:wfj}
\end{equation}
$\mathbf{f}_{k}$ is a $L_{B}\times1$ vector of the \emph{k}th user;
$\mathbf{v}_{k}$ is the $m_{r}n_{r}\times1$ receive analog beamforming
vector applied by user \emph{$k$}; let $\mathbf{v_{\mathit{k}}\left(\mathit{i}\right)}$
denote the $i_{th}$ element of $\mathbf{v_{\mathit{k}}}$, $|\mathbf{v}_{\mathit{k}}\left(\mathit{i}\right)|=\frac{1}{\sqrt{m_{r}n_{r}}}$
, $i=1,...,m_{r}n_{r}$; $\mathbf{n}_{k}$ is the $m_{r}n_{r}\times1$
Gaussian noise vector at user $k$, i.e., $\mathbf{n}_{k}\sim\mathscr{\mathcal{N}}(\mathbf{0},\sigma_{k}^{2}\mathbf{I})$,
$k=1\text{,...,~\emph{K}}$. 

(2) \textbf{\emph{The STAR-aided path}}

The STAR consists of a sub-wavelength UPA having $\varUpsilon$ passive
elements. According to the subarray structure of the BS, we also partition
the STAR into $L_{s}=M_{s}\times N_{s}$ subSTARs. The $i$th subSTAR
consists of $m_{s,i}\times n_{s,i}$ elements, $i=1,...,L_{s}$. Each
subSTAR is the dual counterpart of every subarray of the BS. Thus,
we have $\varUpsilon=\stackrel[i=1]{L_{s}}{\sum}m_{s,i}n_{s,i}$.
The adjacent elements are separated by $\varrho\geq\lambda_{spp}$,
where $\lambda_{spp}$ is the \emph{Surface Plasmon Polariton} (SPP)
wavelength \cite{sarieddeen_terahertz-band_2019}, which is much smaller
than $\lambda$, the free-space wavelength. For simplicity and without
loss of generality, we let $m_{s,i}=m_{s}$, $n_{s,i}=n_{s},$ $i=1,...,L_{s}$.

Let the THz channels spanning from the BS to STAR, and from the STAR
to user $k$, be denoted by $\mathbf{G}$ and $\mathbf{H}_{k,o}$,
respectively. The STAR transmission/reflection matrix is denoted by
$\mathbf{O^{\mathit{TR}}}$ and the transmission/reflection matrix
from the $k$th subSTAR to user $k$ is denoted by $\mathbf{O}_{k}^{TR}$.
Whether the STAR works in the transmission or reflection mode depends
on which side of the STAR the user is on. For the subSTAR structure,
$\mathbf{\mathbf{O^{\mathit{TR}}}}$ is an $L_{s}m_{s}n_{s}\times L_{s}m_{s}n_{s}$
block matrix

\begin{equation}
\mathbf{\mathbf{O^{\mathit{TR}}}}=\left[\begin{array}{cccc}
\mathbf{O}_{1}^{TR} & \mathbf{0} & \cdots & \mathbf{0}\\
\mathbf{0} & \mathbf{\mathbf{O}}_{2}^{TR} & \cdots & \mathbf{0}\\
\vdots & \vdots & \vdots & \vdots\\
\mathbf{0} & \mathbf{0} & \cdots & \mathbf{\mathbf{O}}_{L_{s}}^{TR}
\end{array}\right],
\end{equation}
where $\mathbf{O}_{k}^{TR}$ is an $m_{s}n_{s}\times m_{s}n_{s}$
matrix; $\mathbf{O}_{k}^{TR}=\mathrm{diag}(\mathbf{q}_{k}^{TR})$,
where $\mathbf{q}_{k}^{TR}=\left[e^{j\theta_{1,k}^{TR}},\ldots,e^{j\theta_{i,k}^{TR}},\ldots,e^{j\theta_{m_{s}n_{s},k}^{TR}}\right]^{T}$,
$\theta_{i,k}^{TR}$ is the phase of the $i$-th transmission/reflection
element in the \emph{k}-th subSTAR, $i=1,...,m_{s}n_{s}$. Without
loss of generality, we assume that the number of subSTARs $L_{s}=K$
and the \emph{k}-th subSTAR transmits/reflects the signals for the
\emph{k}-th user. Thus, we propose a joint hybrid 3D beamforming\textendash BS
array-of-subarray and the corresponding subSTAR architecture, the
implementation details will be given in the following sections. For
each subSTAR, either transmission or reflection occurs.

The received signal of user $k$ can be expressed as

\begin{equation}
\tilde{y}_{k}=\mathbf{v}_{k}^{H}\mathbf{H}_{k,o}\mathbf{O^{\mathit{TR}}}\mathbf{G}\mathbf{WFs}+\mathbf{v}_{k}^{H}\mathbf{n}_{k},\label{eq:via ris}
\end{equation}
where $\mathbf{G}$ is the $L_{s}m_{s}n_{s}\times L_{B}m_{t}n_{t}$
element channel matrix of the line spanning from the BS to the STAR.
Furthermore, $\mathbf{H}_{k,o}$ is the $m_{r}n_{r}\times L_{s}m_{s}n_{s}$
channel matrix of the line emerging from the\emph{ }STAR to the user\emph{
k. }Observe that $\tilde{y}_{k}$ can also be expressed in the form
of the desired signal and interference terms as follows:
\begin{equation}
\tilde{y}_{k}=\mathbf{v}_{k}^{H}\mathbf{H}_{k,o}\mathbf{\mathbf{O^{\mathit{TR}}}}\mathbf{G}\mathbf{W}\mathbf{f}_{k}s_{k}+\mathbf{v}_{k}^{H}\mathbf{H}_{k,o}\mathbf{\mathbf{O^{\mathit{TR}}}}\mathbf{G}\sum_{i\neq k}^{K}\mathbf{\mathbf{W}}\mathbf{f}_{i}s_{i}+\mathbf{v}_{k}^{H}\mathbf{n}_{k}.\label{eq:received signal}
\end{equation}

The STAR-aided THz channel is dominated by the LOS path and some indirect
BS-STAR rays from NLOS propagation due to reflection and scattering
\cite{lin_adaptive_2015} due to the quasi-optical characteristic
of THz signals. Let $\mathbf{G_{\mathit{k}}}$ denote the $m_{t}n_{t}\times m_{s}n_{s}$
channel matrix of the line spanning from the \emph{k}-th BS subarray
to the \emph{k}-th subSTAR, while $\mathbf{H_{\mathit{k}}}$ denote
the $m_{r}n_{r}\times m_{s}n_{s}$ channel matrix of the line spanning
from the \emph{k}-th subSTAR to the \emph{k}-th user. Thus, $\mathbf{G_{\mathit{k}}}$
and $\mathbf{H_{\mathit{k}}}$ are modeled by 

\begin{align}
\mathbf{G_{\mathit{k}}} & =\bar{\mathbf{G}}_{\mathit{k}}+\mathbf{\tilde{G}_{\mathit{k}}}\nonumber \\
 & =\sqrt{m_{t}n_{t}m_{s}n_{s}}[\beta_{1,k}^{L}\mathbf{a}_{sa,k}\left(\delta_{k},\kappa_{k}\right)\mathbf{a}_{t,k}^{H}\left(\psi_{k},\tau_{k}\right)\nonumber \\
 & +\stackrel[i=1]{n_{NL}}{\sum}\beta_{1,k,i}^{NL}\mathbf{a}_{sa,k,i}\left(\delta_{k,i}^{NL},\kappa_{k,i}^{NL}\right)\mathbf{a}_{t,k,i}^{H}\left(\psi_{k,i}^{NL},\tau_{k,i}^{NL}\right)],\label{eq:G}
\end{align}

\begin{align}
\mathbf{H}_{k} & =\mathbf{\bar{H}_{\mathit{k}}}+\mathbf{\tilde{H}_{\mathit{k}}}\nonumber \\
 & =\sqrt{m_{r}n_{r}m_{s}n_{s}}[\beta_{2,k}^{L}\mathbf{a}_{r,k}\left(\vartheta_{k},\phi_{k}\right)\mathbf{a}_{sd,k}^{H}\left(\varsigma_{k},\varphi_{k}\right)\nonumber \\
 & +\stackrel[i=1]{\widetilde{n}_{NL}}{\sum}\beta_{2,k,i}^{NL}\mathbf{a}_{r,k,i}\left(\vartheta_{k,i}^{NL},\phi_{k,i}^{NL}\right)\mathbf{a}_{sd,k,i}^{H}\left(\varsigma_{k,i}^{NL},\varphi_{k,i}^{NL}\right)].\label{eq:arj}
\end{align}
where $\bar{\mathbf{G}}_{\mathit{k}}$ and\emph{ $\mathbf{\bar{H}_{\mathit{k}}}$
}represent the\emph{ }LOS components, while $\mathbf{\tilde{G}_{\mathit{k}}}$\emph{
}and $\mathbf{\tilde{H}_{\mathit{k}}}$ denote the NLOS components;
$\psi_{k}$ ($\psi_{k,i}^{NL}$), $\tau_{k}$ ($\tau_{k,i}^{NL})$
are the azimuth and elevation AODs (\emph{angle of departure}) at
the $k$-th BS subarray, respectively; $\vartheta_{k}$ ($\vartheta_{k,i}^{NL}$),
$\phi_{k}$ ($\phi_{k,i}^{NL}$) are the azimuth and elevation AOAs
(\emph{angle of arrival}) at the $k$-th user, respectively; $\delta_{k}$
($\delta_{k,i}^{NL}$), $\kappa_{k}$$(\kappa_{k,i}^{NL})$ are the
azimuth and elevation AOAs from the $k$-th BS subarray to the $k$-th
subSTAR, respectively; $\varsigma_{k}$($\varsigma_{k,i}^{NL}$),
$\varphi_{k}$ ($\varphi_{k,i}^{NL}$) are the azimuth and elevation
AODs from the $k$-th subSTAR to the $k$-th user, respectively; $n_{NL}$
and $\widetilde{n}_{NL}$ are the numbers of NLOS components; $\beta_{1,k}^{L}$
and $\beta_{2,k}^{L}$ denote the corresponding THz LOS complex gains
given by

\begin{align}
|\beta_{g,k}^{L}|^{2} & =\xi_{g,k}^{L}\left(d_{g,k},f_{h}\right)\nonumber \\
 & =\frac{c^{2}}{\left(4\pi d_{g,k}f_{h}\right)^{2}}\exp\left[-\mu\left(f_{h}\right)d_{g,k}\right],~g=1,2,\label{eq:el-1}
\end{align}
where $\xi_{1,k}^{L}$ and $\xi_{2,k}^{L}$ are the corresponding
THz LOS path-loss, $\mu\left(f_{h}\right)$ is the absorption coefficient
at frequency $f_{h}$, $d_{1,k}$ is the distance from the BS to the
STAR, $d_{2,k}$ is the distance from the STAR to the user, both for
user $k$, and $c$ is the speed of light. Still referring to (\ref{eq:G})
and (\ref{eq:arj}), $\beta_{1,k,i}^{NL}$ and $\beta_{2,k,i}^{NL}$
denote the corresponding THz NLOS complex gains.

In Eq. (\ref{eq:G}) and (\ref{eq:arj}), $\mathbf{a}_{t,k}\left(\psi_{k},\tau_{k}\right)$,
and $\mathbf{a}_{r,k}\left(\vartheta_{k},\phi_{k}\right)$ are the
antenna array steering vectors at the \emph{k}-th BS subarray and
$k$-th user, respectively:

\begin{equation}
\begin{aligned}\mathbf{a}_{t,k}\left(\psi_{k},\tau_{k}\right)\\
= & \frac{1}{\sqrt{m_{t}n_{t}}}\left[1,\ldots,e^{j\frac{2\pi r}{\lambda}[x_{1}\cos\psi_{k}\sin\tau_{k}+y_{1}\sin\psi_{k}\sin\tau_{k}]}\right.\\
 & \left.\ldots,e^{j\frac{2\pi r}{\lambda}[(m_{t}-1)\cos\psi_{k}\sin\tau_{k}+(n_{t}-1)\sin\psi_{k}\sin\tau_{k}]}\right]^{T},
\end{aligned}
\label{eq:atj}
\end{equation}
where $x_{1}$ and $y_{1}$ denote the index of the BS antenna element,
$0<x_{1}\leq m_{t}-1$, $0<y_{1}\leq n_{t}-1$. Additionally, $r$
is the distance between the BS antenna elements, and $\lambda$ represents
the wavelength of THz signals. Still referring to (\ref{eq:arj}),
we have

\begin{equation}
\begin{aligned}\mathbf{a}_{r,k}\left(\vartheta_{k},\phi_{k}\right)\\
= & \frac{1}{\sqrt{m_{r}n_{r}}}\left[1,\ldots,e^{j\frac{2\pi\gamma}{\lambda}[x_{2}\cos\vartheta_{k}\sin\phi_{k}+y_{2}\sin\vartheta_{k}\sin\phi_{k}]}\right.\\
 & \left.\ldots,e^{j\frac{2\pi\gamma}{\lambda}[(m_{r}-1)\cos\vartheta_{k}\sin\phi_{k}+(n_{r}-1)\sin\vartheta_{k}\sin\phi_{k}]}\right]^{T},
\end{aligned}
\label{eq:arj-13}
\end{equation}
where $x_{2}$ and $y_{2}$ denote the index of the user antenna element,
$0<x_{2}\leq m_{r}-1$, $0<y_{2}\leq n_{r}-1$; and $\gamma$ is the
distance between the user antenna elements.

Explicitly, $\mathbf{a}_{sa,k}(\delta_{k},\kappa_{k})$ and $\mathbf{a}_{sd,k}(\varsigma_{k},\varphi_{k})$
in Eq. (\ref{eq:G}) and (\ref{eq:arj}) are the arrival and departure
steering vectors at the $k$-th subSTAR, respectively. They can be
expressed as follows:

\begin{equation}
\begin{aligned}\mathbf{a}_{sa,k}(\delta_{k},\kappa_{k})\\
= & \frac{1}{\sqrt{m_{s}n_{s}}}\left[1,\ldots,e^{j\frac{2\pi\varrho}{\lambda}[x_{s}\cos\delta_{k}\sin\kappa_{k}+y_{s}\sin\delta_{k}\sin\kappa_{k}]}\right.\\
 & \left.\ldots,e^{j\frac{2\pi\varrho}{\lambda}[(m_{s}-1)\cos\delta_{k}\sin\kappa_{k}+(n_{s}-1)\sin\delta_{k}\sin\kappa_{k}]}\right]^{T},
\end{aligned}
\label{eq:asaj}
\end{equation}

\begin{equation}
\begin{aligned}\mathbf{a}_{sd,k}(\varsigma_{k},\varphi_{k})\\
= & \frac{1}{\sqrt{m_{s}n_{s}}}\left[1,\ldots,e^{j\frac{2\pi\varrho}{\lambda}[x_{s}\cos\varsigma_{k}\sin\varphi_{k}+y_{s}\sin\varsigma_{k}\sin\varphi_{k}]}\right.\\
 & \left.\ldots,e^{j\frac{2\pi\varrho}{\lambda}[(m_{s}-1)\cos\varsigma_{k}\sin\varphi_{k}+(n_{s}-1)\sin\varsigma_{k}\sin\varphi_{k}]}\right]^{T},
\end{aligned}
\label{eq:asdj}
\end{equation}
where $x_{s}$ and $y_{s}$ denote the index of the STAR element,
$0<x_{s}\leq m_{s}-1$, $0<y_{s}\leq n_{s}-1$; and $\varrho$ is
the distance between the STAR elements within the subSTAR.

Therefore, upon combining (\ref{eq:direct}) and (\ref{eq:via ris}),
the signal received by user $k$ from the BS-user and from the BS-STAR-user
channels can be expressed as

\begin{equation}
y_{k}=\bar{y}_{k}+\tilde{y}_{k}.\label{eq:rec sum}
\end{equation}

The THz channels are sparse and only few paths exists \cite{lin_adaptive_2015}.
Moreover, the power difference of the THz signals between the LOS
and NLOS path is significant. Specifically, the power of the first-order
reflected path is attenuated by more than 10 dB on average compared
to the LOS path and that of the second-order reflection by more than
20 dB \cite{lin_adaptive_2015}, so THz channels are LOS-dominant.
As such, we will focus on the LOS path of the THz signal when exploring
the spatial multiplexing condition and optimal position of the UAV-mounted
STAR. Furthermore, considering the beamforming gain of the transceivers
and the gains of the STAR, the received power of the BS-user link
in (\ref{eq:rec sum}) is much lower than that of the BS-STAR-user
link.

\subsection{Problem Formulation}

The problem is formulated as joint design of the THz BS and STAR (with
the optimal position) with 3D beamforming (both active and passive)
to achieve the optimal performance of multi-user THz massive MIMO
systems for diverse communication environments with different obstacles.

The system performance in terms of the achievable sum-rate of the
users can be expressed by

\begin{equation}
R=\mathbb{E}\left\{ \stackrel[k=1]{K}{\sum}r_{k}\right\} ,
\end{equation}
where we have

\begin{equation}
r_{k}=\log_{2}\left(1+\frac{\frac{P}{K}\left|\mathbf{v}_{k}^{H}\mathbf{H}_{k}\mathbf{O}^{TR}\mathbf{G}\mathbf{\mathbf{\mathbf{W}}}\mathbf{f}_{k}\right|^{2}}{\frac{P}{K}\sum_{i\neq k}^{K}\left|\mathbf{v}_{k}^{H}\mathbf{H}_{k}\mathbf{O}^{TR}\mathbf{\mathbf{G}W}\mathbf{f}_{i}\right|^{2}+\left|\mathbf{v}_{k}^{H}\mathbf{n}_{k}\right|^{2}}\right).
\end{equation}

With GI quantum sensing, we can obtain the information of the communication
environment including the obstacles and positions of the potential
users. The details will be given in the following sections. The 3D
image of the environment obtained from GI is discretized due to limited
resolutions, which can be easily transformed into 3D grid $\mathcal{\mathcal{F}}$
including vertices and edges, where $\mathcal{\mathcal{F}}$\emph{
}belongs to\emph{ }the Euclidean space\emph{ }$\mathbb{R}^{3}$. The
grid granularity is approximately equal to the resolution of GI 3D
imaging. Then we perform K-means to obtain the user clustering and
centroids of the user clusters. For simplicity and without loss of
generality, we focus on one user cluster here. The 3D positions of
the BS and centroid of the user cluster, denoted by $\mathbf{b}_{\mathit{BS}}$$\left(x_{B},y_{B},z_{B}\right)$
and $\mathbf{b}_{\mathit{user}}\left(x_{u},y_{u},z_{u}\right)$, respectively,
are transformed into vertices of $\mathcal{\mathcal{F}}$. Let $\mathbf{b}_{\mathit{STAR}}$$\left(x_{A},y_{A},z_{A}\right)$
denote the position of the UAV-mounted STAR. The set of obstacles
is represented by $\varXi$. 

Our objective is to maximize the sum rate of all the users. The optimization
problem is formulated as

\begin{align}
 & \mathrm{\underset{\{\mathbf{v},\mathbf{O}^{\mathit{TR}},\mathbf{W},\mathbf{f},\mathbf{b}_{\mathit{STAR}}\}}{maximize}}R,\nonumber \\
\mathrm{s.t}.~~ & \mathrm{C}1:|[\mathbf{\mathbf{O^{\mathit{TR}}}}]_{n,n}|=1.\nonumber \\
 & \mathrm{C}2:\varXi\subseteq\mathcal{\mathcal{F}},\mathbf{b}_{\mathit{BS}},\mathbf{b}_{\mathit{user}},\mathbf{b}_{\mathit{STAR}}\in\mathcal{\mathcal{F}};\nonumber \\
 & \mathrm{C}3:||\mathbf{b}_{\mathit{BS}}-\mathbf{b}_{\mathit{STAR}}||+||\mathbf{b}_{\mathit{STAR}}-\mathbf{b}_{\mathit{user}}||>\nonumber \\
 & ~~||\mathbf{b}_{\mathit{BS}}-\mathbf{b}_{\mathit{user}}||;\nonumber \\
 & \mathrm{C}4:\Psi_{1}=\{\mathbf{b}\mid||\mathbf{b}_{\mathit{BS}}-\mathbf{b}||+||\mathbf{b}-\mathbf{b}_{\mathit{STAR}}||\nonumber \\
 & ~~=||\mathbf{b}_{\mathit{BS}}-\mathbf{b}_{\mathit{STAR}}||,\mathbf{b}\in\mathcal{\mathcal{F}}\};\nonumber \\
 & \mathrm{C}5:\Psi_{2}=\{\mathbf{b}\mid||\mathbf{b}_{\mathit{STAR}}-\mathbf{b}||+||\mathbf{b}-\mathbf{b}_{\mathit{user}}||\nonumber \\
 & ~~=||\mathbf{b}_{\mathit{STAR}}-\mathbf{b}_{\mathit{user}}||,\mathbf{b}\in\mathcal{\mathcal{F}}\};\nonumber \\
 & \mathrm{C}6:\Psi\cap\varXi=\phi,~where~\Psi=\Psi_{1}\cup\Psi_{2};\nonumber \\
 & \mathrm{C7:}\tilde{\Psi}\cap\varXi\neq\phi,~where~\tilde{\Psi}=\nonumber \\
 & ~~\{\mathbf{b}\mid||\mathbf{b}_{\mathit{BS}}-\mathbf{b}||+||\mathbf{b}-\mathbf{b}_{\mathit{user}}||=\nonumber \\
 & ~~||\mathbf{b}_{\mathit{\mathit{BS}}}-\mathbf{b}_{\mathit{\mathit{user}}}||,\mathbf{b}\in\mathcal{\mathcal{F}}\};\nonumber \\
 & \mathrm{C}8:||\mathbf{W}\mathbf{f}_{k}||^{2}=1,~k=1...K;\nonumber \\
 & \mathrm{C}9:z_{min}\leq z_{A}\leq z_{max}.\label{eq:OBJ}
\end{align}
Constraint C1 represents the STAR unit-modulus constraint of phase
shift matrices; Constraint C2 restricts that the positions of the
BS, user and STAR should be inside the considered environment $\mathcal{\mathcal{F}}$;
Constraint C3 restricts that the sum LOS distance from the BS to the
STAR and from the STAR to the user should be larger than the direct
distance between the BS and the user; Constraint C4, and C5 restricts
that \emph{$\Psi_{1}$},\emph{ $\Psi_{2}$} are the LOS path from\emph{
$\mathbf{b}_{\mathit{BS}}$ }to\emph{ $\mathbf{b}_{\mathit{STAR}}$},
and from\emph{ $\mathbf{b}_{\mathit{STAR}}$ }to\emph{ $\mathbf{b}_{\mathit{user}}$},
respectively; Constraint C6 restricts that the cascaded LOS path $\Psi$
(where paths \emph{$\Psi_{1}$ }and\emph{ $\Psi_{2}$} are connected)
assisted by UAV-mounted STAR should not intersect the interior of
any obstacle\emph{ $\varXi$}; Constraint C7 restricts that the LOS
path $\tilde{\Psi}$ from the BS to the user is obstructed by the
obstacles $\varXi$; Constraint C8 represents the conditions that
the analog transmit beamforming and baseband digital beamforming needs
to satisfy; Constraint C9 restricts the the minimum flight altitude
$z_{min}$ and the maximum flight altitude $z_{max}$ of the UAV which
may be imposed by government regulations.

\section{Conditions of Spatial Multiplexing for the Proposed Architecture}

In this section, we discuss the conditions of achieving high multiplexing
gains for the proposed STARs-aided THz architecture. As the propagation
of signals at THz frequencies is ``quasi-optical'', the LOS path
dominates the channel complemented only by a few non-LOS (NLOS) reflected
rays due to the associated high reflection loss. Thanks to the beamforming
gain and flexible placement of STARs, a LOS path may be present between
each pair of the BS subarrays, the subSTAR and the user's receiver
arrays. 

We first consider the BS-STAR channel. Without loss of generality,
we assume symmetry in the remainder of the paper, i.e., $M_{t}=N_{t}=M$,
$M_{s}=N_{s}=N$. Thus, the BS contains $M\times M$ subarrays and
the STAR contains $N\times N$ subSTARs. The capacity is maximized
when all columns of $\mathbf{G}$ are orthogonal. Because STARs can
transmit and reflect at the same time, they can provide omnidirectional
coverage compared to reflection-only RISs, and thus a STAR can simply
be placed in front of BS without an angle. Based on \cite{wang_joint_2022},
we have the following Lemma for STAR-aided signal transmission/reflections. 
\begin{lem}
For STAR-aided systems, the optimal subSTARs and BS subarray design
condition should satisfy 

\begin{equation}
\alpha_{h}=\chi_{h}=\sqrt{q\frac{cd_{1}}{f_{h}}\Delta},
\end{equation}
for integer values of $q$, and $\alpha_{h}\ll\lambda$, $\chi_{h}\ll\lambda$,
where $\Delta$ is the minimal common multiples of $\left\{ \frac{1}{M},\frac{1}{N}\right\} $\emph{;}
$\alpha_{h}$ and $\chi_{h}$\emph{ are }the optimal distances between
two adjacent BS subarrays and subSTARs for the $h$th THz subband\emph{,
respectively; $d_{1}$ }is the distance between the centers of the
STAR and BS arrays.
\end{lem}
\begin{IEEEproof}
The proof is similar to Theorem 1 in \cite{wang_joint_2022}, hence
it is omitted here. 
\end{IEEEproof}
The separation of elements in the BS subarrays or subSTARs may be
achieved via spatial interleaving. Moreover, the required spacing
can be realized by choosing the right elements belonging to each BS
subarray or subSTAR. Each BS subarray associated with the corresponding
subSTAR may focus on a specific individual subband of the THz signal
and can also be tuned flexibly according to the different user distances. 

Let us now consider the channels spanning from the subSTARs to the
users. At the receiver side, since each user has an antenna array
and is generally separated, the distance between two adjacent user
antenn arrays is usually large. Naturally, we cannot impose any constraints
on the user positions, which tend to be random. Therefore, we must
resort to exploiting the frequency selectivity of the THz channels
for spatial multiplexing. According to \cite{lin_adaptive_2015},
the THz channels have limited angular spread of about $40\lyxmathsym{\textdegree}$.
As the coverage area of a STAR is greatly expanded to almost 360 degrees,
the requirement of such spatial independence becomes much easier to
satisfy than a reflection-only RIS. Since the beam steering vectors
associated with completely different angles of large-scale antennas
are nearly orthogonal, for the channels spanning from the \emph{i}th
and the \emph{k}th subSTAR to user\emph{ k, }we have

\begin{equation}
\mathbf{a}_{sd,i}^{H}(\varsigma_{i},\varphi_{i})\mathbf{a}_{sd,k}(\varsigma_{k},\varphi_{k})\simeq0,\quad\varsigma_{i}\neq\varsigma_{k},\varphi_{i}\neq\varphi_{k}.\label{eq:angle}
\end{equation}
As the STAR may be regarded as a passive large-scale antenna array,
the channel from the STAR to the user can be nearly orthogonal, hence
spatial multiplexing gains can be achieved. For users that are close
enough to be within the angular spread, the pre-scanning and grouping
technique of \cite{lin_adaptive_2015} can be adopted.

\section{Proposed Quantum Sensing based Joint 3D Beam Training for the UAV-mounted
STAR aided THz multi-user MIMO Systems}

\begin{figure}[htbp]
\begin{centering}
\includegraphics[scale=0.5]{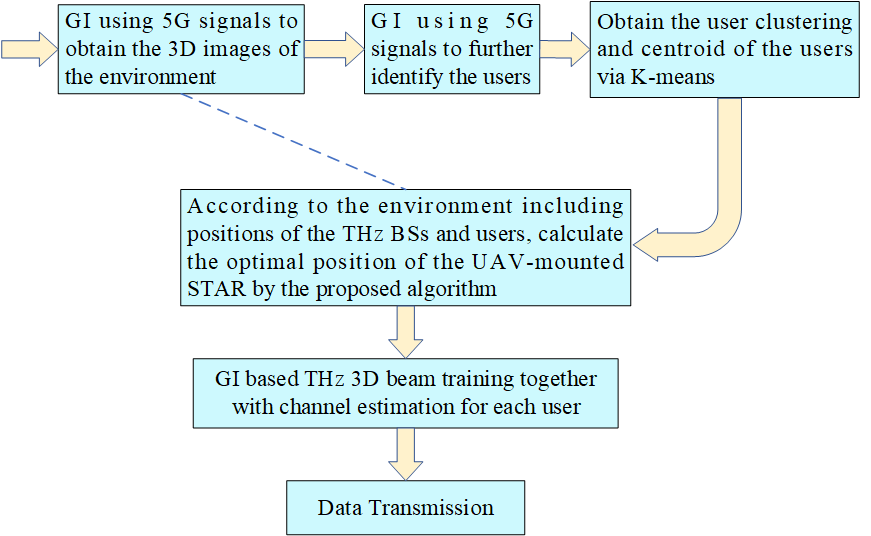}
\par\end{centering}
\caption{\label{fig:Block-diagram-of}Flow diagram of the proposed Ghost Imaging
based joint 3D beam training for the UAV-mounted STAR aided THz multi-user
MIMO systems }
\end{figure}

In order to realize the proposed joint hybrid 3D beamforming BS array-of-subarray
and the corresponding subSTAR architecture as well as the effective
communication environment sensing, in this section, we propose a novel
quantum sensing, i.e., Ghost Imaging based beam training scheme for
UAV-mounted STAR aided THz multi-user MIMO systems, the flow diagram
of which is shown in Fig. \ref{fig:Block-diagram-of}. The environment
information obtained by GI allows us to expedite the channel estimation
and beamforming processes. We first conduct GI using 5G BSs to obtain
the 3D images of the environment and users. Then we perform K-means
to obtain the user clustering and centroid of the users. For simplicity
and without loss of generality, we focus on one user cluster here.
According to the environment including positions of the THz BSs and
users, the optimal position of the UAV-mounted STAR can be calculated
by the proposed algorithm. We also propose a semi-passive structure
of the STAR for channel estimation. Based on the estimated separated
channel information, active and passive beamforming are carried out
and data transmission begins. We will explain in detail each block
function in the following subsections. 

\subsection{Ghost Imaging by surrounding 5G BSs}

\begin{figure}[tbh]
\includegraphics[scale=0.35]{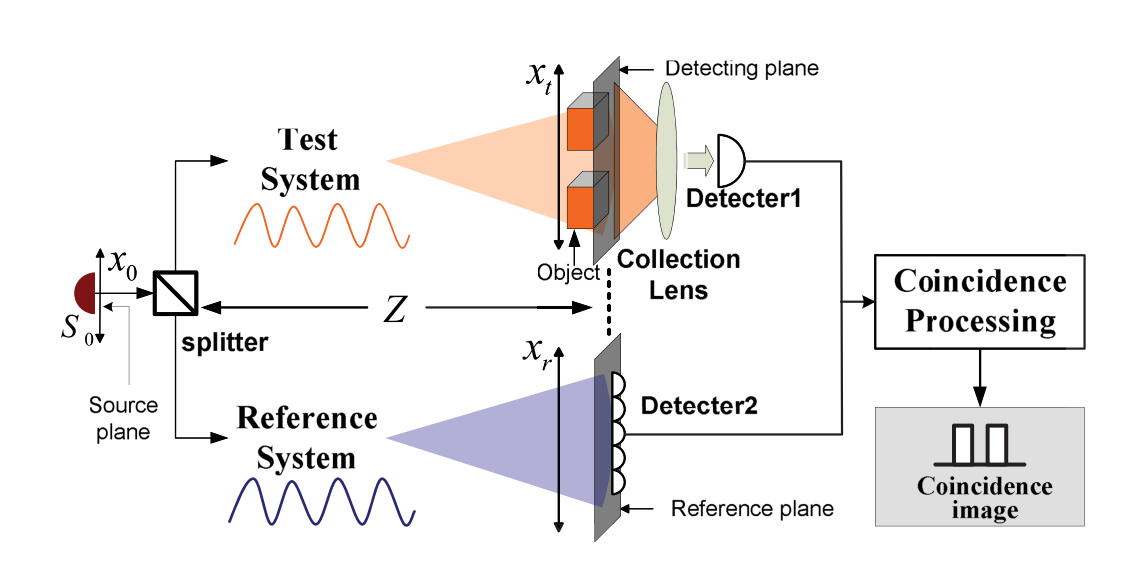}\caption{\label{fig:Classical}Classical coincidence imaging \cite{dongze_li_radar_2014}.}
\end{figure}

Multiple communication systems must coexist in future communication
scenarios in which the near-ubiquitous surrounding 5G BSs can be used
to assist in Ghost imaging for low-cost ISAC systems. Thus we first
conduct GI by surrounding 5G downlink signals to obtain the 3D images
of the environment including users and obstacles. Inspired by the
classical coincidence imaging implemented in optical system as shown
in Fig. \ref{fig:Classical}, an instantaneous microwave radar imaging
technique called \emph{Radar Coincidence Imaging} (RCI) \cite{dongze_li_radar_2014}
can obtain high-resolution images of target focus without motion limitation.
 The high resolution reconstruction of the target can be achieved
in the GI framework by mimicking the radiation element on the surface
of the incoherent light source with an antenna emitting randomly modulated
signals. Microwave GI or association imaging is to generate radiation
field distribution with time-varying and space-varying characteristics
in the microwave frequency band, form differential irradiation on
the imaging area, associate the target echo signal with the radiation
field, and reconstruct the target image. Because the resolution of
the target is derived from the modulation of the transmitted signal,
it breaks the dependence of the traditional microwave imaging technology
on the observation angle and Doppler, and does not need the relative
motion between the imaging system and the target.

Microwave GI can be modeled as a linear model, and imaging reconstruction
can be regarded as solving a linear inverse problem. According to
the radiation field reference matrix and the properties of the target
scattering coefficient vector, multiple methods can be used to obtain
the estimation of the target scattering coefficient vector, so as
to reconstruct the target image. First, consider the EM field on plane
\emph{$\varOmega$} radiated by $N_{L}$ 5G BSs, then we have

\begin{equation}
E(\Delta x,t)=\sum_{i=1}^{N_{L}}S_{i}^{L}\left(t-\tau_{i,\Delta x}\right)\ell_{i,\Delta x},
\end{equation}
where $E(\Delta x,t)$ is the EM state of the single point $\Delta x$
on the imaging plane $\varOmega$, $\Delta x\in\varOmega$, \emph{$\tau_{i,\Delta x}$}
is the propagation delay from the \emph{i}th BS to $\Delta x$, and
$\ell_{i,\Delta x}$ is the propagation attenuation. The randomness
of the propagation delays from these non-uniformly distributed BSs
further enhances the spatial incoherence of the generated EM field.

To reconstruct the image of objects, the sampling interval should
be set much larger than the coherence time to guarantee the incoherence
between any two samples. That is, $T_{s}>\triangle t$, where $\triangle t=\frac{1}{W_{L}}$,
$W_{L}$ is the bandwidth of the 5G signals. 

According to Born's Approximation, we divide the imaging area \emph{$\varOmega$}
equally into \emph{$M$} sub-planes with \emph{$I$} rows and \emph{$J$}
columns, where $M=IJ$. It is worth mentioning that although we consider
two-dimensional image planes here, the technique of tomographic imaging
can be used to construct three-dimensional images.

The illumination matrix on the surface of the target plane after \emph{$N_{m}$}
transmissions from 5G BSs can be expressed by

\begin{equation}
\mathbf{E}=\left[\begin{array}{cccc}
E_{1,1}^{1} & E_{2,1}^{1} & \ldots & E_{I,J}^{1}\\
E_{1,1}^{2} & E_{2,1}^{2} & \ldots & E_{I,J}^{2}\\
\vdots & \vdots & \ddots & \vdots\\
E_{1,1}^{N_{m}} & E_{2,1}^{N_{m}} & \ldots & E_{I,J}^{N_{m}}
\end{array}\right],
\end{equation}
where $E_{i,j}^{n}$, $n=1,...,N_{m},i=1,...,I,j=1,...,J$, is the
EM field at the corresponding pixel in the $n$th measurement.

Then the received signal can be expressed as 

\begin{equation}
\mathbf{y}=\mathbf{E}\boldsymbol{\epsilon}\mathbf{\boldsymbol{\rho}},
\end{equation}
where 

\begin{equation}
\epsilon=\left[\epsilon_{1,1},\epsilon_{2,1},\ldots,\epsilon_{I,J}\right]^{T},
\end{equation}

\begin{equation}
\boldsymbol{\mathbf{\rho}}=\mathrm{diag}\left[\rho_{1,1},\rho_{2,1},\ldots,\rho_{I,J}\right],
\end{equation}

\begin{equation}
\mathbf{y}=\left[y_{1},y_{2},\ldots,y_{N}\right]^{T},
\end{equation}
$\epsilon_{p,q}$ is the corresponding scattering coefficient, $y_{n}$
is the receiving signal from the $n$th illumination and $\rho_{p,q}$
is the propagation attenuation from the pixel to the receiver. The
image reconstruction of the object can be achieved by solving the
following optimization problem

\begin{equation}
\mathrm{\underset{\boldsymbol{\sigma}}{\mathit{arg}\ min}}\left\{ \left\Vert \mathbf{y}-\mathbf{E}\boldsymbol{\epsilon}\mathbf{\boldsymbol{\rho}}\right\Vert ^{2}\right\} .\label{eq:miny}
\end{equation}
Optimization algorithms, e.g., \emph{Gradient Projection} (GP), can
be used to solve it. Here we adopt the \emph{minimum mean squared
error} (MMSE) method. 

The spatial resolution of microwave GI is restricted by the physical
size of the coherent size, the boundary of which in the background
field can be expressed as \cite{wang_microwave_2016}

\begin{equation}
\frac{D_{st}\lambda}{2\Upsilon}\leq d_{c}\leq\frac{D_{st}\lambda}{\Upsilon},
\end{equation}
where $\Upsilon$ is the size of the source, $\lambda$ is the wavelength,
and $D_{st}$ is the normal distance from the source to the target
objects. 

\begin{figure}[tbh]
\begin{centering}
\includegraphics[scale=0.35]{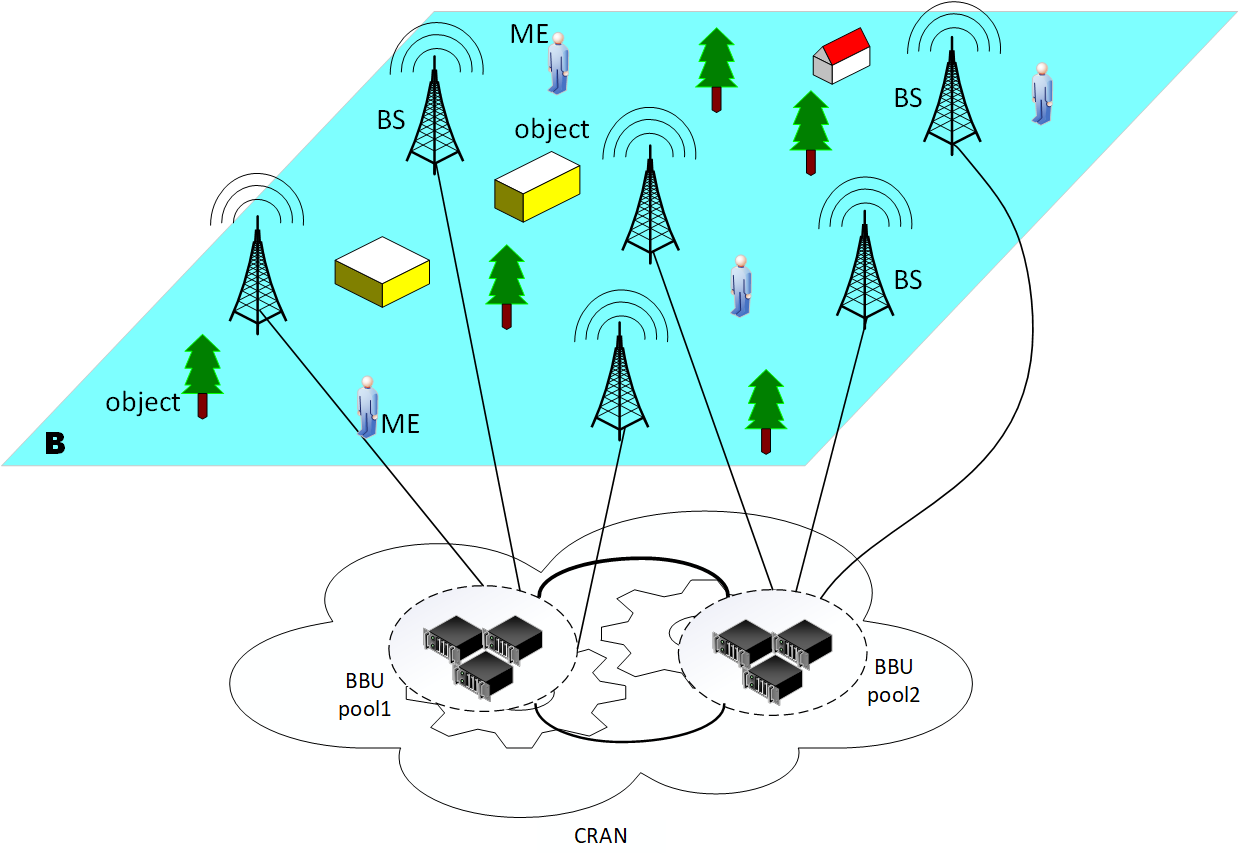}
\par\end{centering}
\caption{\label{fig:System-model-GI}System model of the ghost imaging based
on 5G CRAN.}
\end{figure}

Here we propose to utilize the CRAN structure of the 5G BSs, which
can further enlarge the size of the source. The three-dimensional
(3D) 5G-based GI scenario based on CRAN is depicted in Fig. \ref{fig:System-model-GI}.
The investigation area $\varOmega$ is inhomogeneous with several
objects to be imaged. The objects in $\varOmega$ are distinguished
by their scattering coefficients. The investigation area \emph{$\varOmega$}
is illuminated by 5G BSs. The height from BSs to $\varOmega$ are
assumed to be identical. Let $S_{i}^{L}$ denote the signals transmitted
from the $i$th BS, and reflected signals from objects are collected
by a single receiving BS located in the centre of $\varOmega$. THz
BSs are co-located with 5G BSs, or their core networks can share information.

\begin{figure}[tp]
\begin{centering}
\includegraphics[scale=0.38]{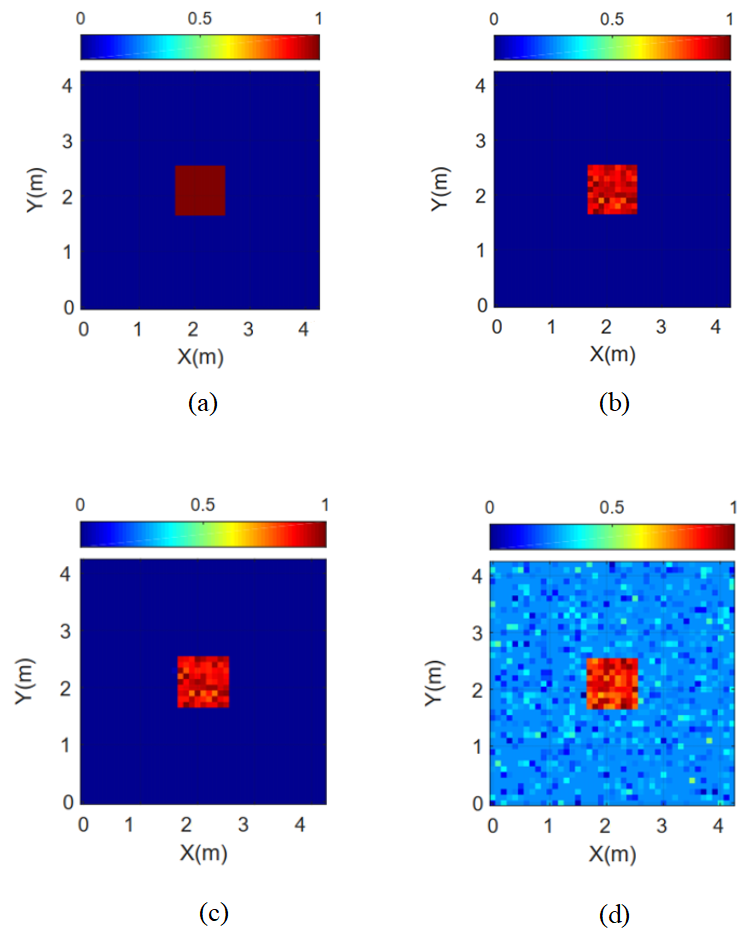}
\par\end{centering}
\caption{\label{fig:GI} GI Reconstruction results with improved resolution
compared to \cite{zhang_microwave_2018}. (a) Original scenario; (b)
SNR = 30 dB; (c) SNR = 20 dB and (d) SNR = 10 dB.}
\end{figure}

Thus the resolution of the 5G GI can be much higher than that of the
traditional LTE GI. The simulation results are shown in Fig. \ref{fig:GI}
with a single square shaped target located in the center of the investigation
area $\varOmega$ as an example. As shown in Fig. \ref{fig:GI} (c),
(d) and (e), the GI reconstruction of the object is achieved with
much improved resolution ($0.1\mathrm{m}\times0.1\mathrm{m}$) compared
to that of the LTE GI \cite{zhang_microwave_2018} ($10\mathrm{m}\times10\mathrm{m}$)
under the different SNR. The edges of the objects or users detected
can be adopted to assist the optimal UAV-mounted STAR position finding
as well as the joint active and passive beam forming.

\subsection{Joint Active and Passive Beamforming Design of the UAV-mounted STAR}

The BS utilizes the angle information of the BS-RIS link to design
the active beamforming. For THz signals, the pencil-like beams necessitate
perfect line-of-sight (LoS) alignment between the transmitter and
receiver antennas. The phase shifts of the signals reflected/transmitted
from STARs can be proactively adjusted to establish virtual LoS links.
Without loss of generality, we assume that the \emph{k}-th user is
assisted by the \emph{k}-th subSTAR. Thus, the active beamforming
designed for the $k$-th user should be aligned to the $k$-th subSTAR.
As such, the transmit beam of the $k$-th BS subarray is designed
as 

\emph{
\begin{equation}
\mathbf{w}_{k}=\sqrt{p_{k}}\mathbf{a}_{t,k}\left(\psi_{k},\sigma_{k}\right),\label{eq:wj}
\end{equation}
}where $p_{k}$ is the transmit power of the \emph{k}-th BS subarray.
When the transmit power is equally divided among BS subarrays corresponding
to different users, we have $p_{k}$=$\frac{P_{s}}{K}$. 

The received beam of the $k$-th user is given by
\begin{equation}
\mathbf{\hat{v}}_{k}=\sqrt{m_{r}n_{r}}\mathbf{a}_{r,k}\left(\hat{\vartheta}_{k},\hat{\phi}_{k}\right).\label{eq:v}
\end{equation}

Based on \cite{wang_joint_2022}, we design the optimal STAR phase
shift beam as 

\begin{equation}
\mathbf{\hat{q}}_{k}^{TR}=\frac{1}{\hat{\beta}_{2,k}^{L}}\left[\left(\hat{\mathbf{\bar{H}}}_{k}^{T}\mathbf{\hat{v}}_{k}^{*}\right)\odot\boldsymbol{\mathrm{a}}_{sa,k}\right]^{*}.\label{eq:w}
\end{equation}
where $\hat{\bar{\mathbf{H}}}_{k}=\sqrt{m_{r}n_{r}m_{s}n_{s}}\hat{\beta}_{2,k}^{L}\mathbf{a}_{r,k}\left(\hat{\vartheta}_{k},\hat{\phi}_{k}\right)\mathbf{a}_{sd,k}^{H}\left(\hat{\varsigma}_{k},\hat{\varphi}_{k}\right)$,
and $\hat{\beta}_{2,k}^{L}$ is the estimation of $\beta_{2,k}^{L}$.

The AOAs and AODs can be calculated from the BS and user positions
obtained by GI sensing, and the optimal position of UAV-mounted STAR
which will be acquired in proposed Algorithm 1 presented later. Then
the pre-specified codebook designed with directional beams associated
with the corresponding AOAs/AODs can be selected, thus GI based beam
training is achieved. 

To further enhance the beamforming gain, the instantaneous channel
state information (CSI) can be obtained by e.g., tensor-based channel
estimation \cite{de_araujo_channel_2021}. All the beam training and
channel estimation can be carried out in parallel for every pair of
BS-subarray, subSTAR and user array. Furthermore, we propose a semi-passive
STAR structure as shown in Fig. \ref{fig:System-model-of} and the
channel estimation ambiguity elimination scheme for separate channel
estimation, the details of which can be found in Appendix A, where
it is shown that the ambiguity in the separated channel\textbf{ $\mathbf{U}^{k}$
}and\textbf{ $\mathbf{\mathbf{\mathbf{A^{\mathit{k}}}}}$ }may affect
the optimal passive beamforming design. Therefore, it can be noted
that the separate channel estimation instead of cascaded channel estimation
is adopted above for the optimal active and passive beamforming design. 

\subsection{Design of the BS's Digital Precoder }

Following the beam training and channel estimation, we will now design
the digital \emph{transmit precoder} (TPC) for interference cancellation
among different users. The digital TPC can be designed as follows.
Let

\begin{equation}
\mathbf{\hat{T}}_{k}=\mathbf{\hat{v}}_{k}^{H}\hat{\bar{\mathbf{H}}}_{k}\mathbf{O^{\mathit{TR}}}\mathbf{\hat{G}}\mathbf{W},
\end{equation}
where $\mathbf{\hat{G}}$ denote the estimated channel of $\mathbf{G}$.
Specifically, the MMSE TPC is formulated as

\begin{equation}
\mathbf{F}=\left[\mathbf{f}_{1},\mathbf{f}_{2},\ldots,\mathbf{f}_{K}\right]=\left[\left(\hat{\mathbf{T}}^{H}\hat{\mathbf{T}}+\frac{K\sigma^{2}}{P}\mathbf{W}^{H}\mathbf{W}\right)^{-1}\hat{\mathbf{T}}^{H}\right]^{-1}.\label{eq:f}
\end{equation}

\subsection{Problem Reformulation and Proposed Solution}

According to the orthogonality of different users arranged by the
subSTAR and subarray of the BS, which is achieved by the conditions
shown in Section III, as well as the previous digital TPC, we have

\begin{equation}
\frac{P}{K}\sum_{i\neq k}^{K}\left|\mathbf{v}_{k}^{H}\hat{\bar{\mathbf{H}}}_{k}\mathbf{O}^{TR}\mathbf{\mathbf{\hat{G}}}\mathbf{\mathbf{\mathbf{W}}}\mathbf{f}_{i}\right|^{2}\approx0.
\end{equation}
Therefore, we arrive at:

\begin{equation}
\hat{r}_{k}=\log_{2}\left(1+\frac{\frac{P}{K}\left|\mathbf{v}_{k}^{H}\hat{\bar{\mathbf{H}}}_{k}\mathbf{O}^{TR}\mathbf{\mathbf{\mathbf{\hat{G}}}}\mathbf{W}\mathbf{f}_{k}\right|^{2}}{\left|\mathbf{v}_{k}^{H}\mathbf{n}_{k}\right|^{2}}\right).\label{eq:rk}
\end{equation}

Then the optimization problem in (\ref{eq:OBJ}) becomes

\begin{align}
\mathrm{\underset{\{\mathbf{v},\mathbf{O}^{\mathit{TR}},\mathbf{W},\mathbf{f},\mathbf{b}_{\mathit{STAR}}\}}{maximize}} & \mathbb{E}\left\{ \stackrel[k=1]{K}{\sum}\hat{r}_{k}\right\} \label{eq:r head}\\
\mathrm{s.t.} & \mathrm{C}1-\mathrm{C}9,
\end{align}
which is a non-convex combinatorial optimization problem which cannot
be solved efficiently via standard optimization methods and the global
optimum is challenging to find. To address (\ref{eq:r head}), we
decouple the optimization into three sub-problems, i.e., the joint
active and passive beamforming, the design of digital precoding, and
position optimization of the UAV-mounted STAR.

Let us represent the \emph{k}th subSTAR-aided cascaded channel by

\begin{equation}
\mathbf{Z}_{k}=\hat{\bar{\mathbf{H}}}_{k}\mathbf{O}\mathbf{\mathbf{\mathbf{\mathbf{\hat{G}}}}}.
\end{equation}

\noindent For far-field beamforming over the STAR-aided THz channel,
the channel power of $\mathbf{Z}_{k}$ of the \emph{k}th user can
be expressed as

\begin{equation}
\hat{\zeta}_{k}=\frac{c^{2}}{\left(4\pi f\right)^{2}d_{1}^{2}d_{2,k}^{2}}e^{-\mu(f)\left(d_{1}+d_{2,k}\right)}.\label{eq:far}
\end{equation}
where $d_{2,k}$ is the distance from the STAR to the user \emph{k}
and $d_{1}=||\mathbf{b}_{\mathit{BS}}-\mathbf{b}_{\mathit{STAR}}||$.
Let $\overline{d}_{2}$ denote the the distance from the STAR to the
centroid of the user cluster, i.e., $\overline{d}_{2}$=$||\mathbf{b}_{\mathit{STAR}}-\mathbf{b}_{\mathit{user}}||$.
Let $D=||\mathbf{b}_{\mathit{BS}}-\mathbf{b}_{\mathit{user}}||$.
Based on the previous obtained optimal active and STAR passive beamforming
in (\ref{eq:wj}), (\ref{eq:v}), (\ref{eq:w}) and (\ref{eq:f})
for $\mathbf{v}$, $\mathbf{O}^{\mathit{TR}}$, $\mathbf{W}$ and
digital precoding $\mathbf{f}$, respectively, we can now opt for
the last sub-problem, i.e., the end-to-end THz LoS channel power optimization
through optimizing the position of the UAV-mounted STAR $\mathbf{b}_{\mathit{STAR}}$:

\begin{align}
\underset{\mathbf{b}_{\mathit{STAR}}}{\mathrm{maximize~}} & \frac{c^{2}}{\left(4\pi f\right)^{2}d_{1}^{2}\overline{d}_{2}^{2}}e^{-\mu(f)\left(d_{1}+\overline{d}_{2}\right)},\nonumber \\
\mathrm{s.t}.~ & \mathrm{C}10:d_{1}+\overline{d}_{2}\geq D\nonumber \\
 & \mathrm{C}2,\mathrm{C}4,\mathrm{C}5-\mathrm{C}9\label{eq:qq1}
\end{align}
By discarding irrelevant constant terms and taking logarithm of (\ref{eq:qq1}),
the optimization problem of (\ref{eq:qq1}) is equivalent to

\begin{align}
\underset{\mathbf{b}_{\mathit{STAR}}}{\mathrm{minimize~}} & \mu(f)\left(d_{1}+\overline{d}_{2}\right)+2\left(\mathrm{ln}d_{1}+\mathrm{ln}\overline{d}_{2}\right),\nonumber \\
\mathrm{s.t.}~ & \mathrm{C}2,\mathrm{C}4,\mathrm{C}5-\mathrm{C}10\label{eq:q2}
\end{align}
As $\mu(f)$ is about $10^{-6}$ \cite{yao_stochastic_2017}, generally
for far-field, $\sqrt{e}\left(\approx1.65\right)<d_{1},~\overline{d}_{2}<10^{3}$
\cite{wang_joint_2022}, thus $\mu(f)\left(d_{1}+\overline{d}_{2}\right)$
is about $10^{-3}$, whereas $2\left(\mathrm{ln}d_{1}+\mathrm{ln}\overline{d}_{2}\right)>1$.
Therefore, we have $\mu(f)\left(d_{1}+\overline{d}_{2}\right)\ll2\left(\mathrm{ln}d_{1}+\mathrm{ln}\overline{d}_{2}\right)$,
which makes (\ref{eq:q2}) equivalent to 

\begin{align}
\underset{\mathbf{b}_{\mathit{STAR}}}{\mathrm{minimize~}} & \left(\mathrm{ln}d_{1}+\mathrm{ln}\overline{d}_{2}\right),\nonumber \\
\mathrm{s.t}.~ & \mathrm{C}2,\mathrm{C}4,\mathrm{C}5-\mathrm{C}10\label{eq:q3}
\end{align}

We then need to find out the optimal position $\mathbf{b}_{\mathit{STAR}}$
of the UAV-mounted STAR to solve the problem (\ref{eq:q3}). Although
the shortest paths in deterministic 2D environments can be found by
performing algorithms such as Dijkstra, finding the shortest paths
in 3D environments with polyhedral obstacles is NP-hard \cite{nash_lazy_2010}.
An \textquotedblleft any-angle\textquotedblright{} find-path algorithm
called Theta{*} can finds shorter paths on square grids than both
A{*} and A{*} PS with a similar runtime \cite{nash_lazy_2010}. Theta{*}
typically finds much shorter paths than that by propagating information
along graph edges without constraining paths to be formed by graph
edges. Lazy Theta{*} \cite{nash_lazy_2010} is a variant of Theta{*}
using lazy evaluation to perform only one line-of-sight check per
expanded vertex with slightly more expanded vertices. Lazy Theta{*}
guarantees the optimal path-finding and performs faster than Theta{*}
on cubic grids, with one order of magnitude fewer LOS checks and no
increase in path length. Considering that \emph{$\Psi_{1}$ }and\emph{
$\Psi_{2}$} are LOS paths and Lazy Theta{*} can detect any optimal
LOS path, we tactfully utilize Lazy Theta{*} to find the optimal position
$\mathbf{b}_{\mathit{STAR}}$ (with the path lengths set to logarithm
of the Euclidean length) under the constraints of $\mathrm{C}2,\mathrm{C}4,\mathrm{C}5-\mathrm{C}10$.
Upon execution, Algorithm 1 will output $\mathbf{b}_{\mathit{STAR}}$,
the optimal position of the UAV mounted STAR. Based on (\ref{eq:q3}),
the proposed UAV-mounted STAR position finding algorithm is proposed
as follows. 

\begin{algorithm}[tbh]
\caption{UAV-mounted STAR optimal position finding algorithm}

1 \textbf{Input }$b_{BS}$, $b_{user}$, $nghbr_{LOS}(b)$

2 \ \ $open:=closed:=\varnothing$; 

3 $\ \ c(b_{BS}):=0$; 

4 $\ \ parent(b_{BS}):=b_{BS}$; 

5 $\ \ open.Insert(b_{BS},c(b_{BS})+h(b_{BS}))$; 

6 \textbf{\ \ while} $open\notin\varnothing$ \textbf{do} 

7 $\ \ \ \ b:=open.Pop()$; 

8 \textbf{\ \ \ \ if} NOT $\mathrm{LOS}(parent(b),b)$ \textbf{then} 

9 $\ \ \ \ \ \ parent(b):=$

$~~~\ \ \!\ \ \ \ argmin_{b'\in nghbr_{LOS}(b)\bigcap closed}(c(b')+\ln l(b',b))$; 

10 $\ \ \ \ \ c(b):=min_{b'\in nghbr_{LOS}(b)\bigcap closed}(c(b')+\ln l(b',b));$

11 \ \ \textbf{\ \ end if}; 

12 \textbf{\ \ \ \ if} $b=b_{user}$ \textbf{then }

13 \ \ \ \ \ \ \textbf{output }the optimal position of the STAR
$parent(b_{user})$;

14 \ \ \ \textbf{\ end if}

15 $\ \ \ \ closed:=closed\bigcup{b}$; 

16 \textbf{\ \ \ \ foreach} $b'\in nghbr_{LOS}(b)$ \textbf{do }

17 \textbf{\ \ \ \ \ \ if} $b'\notin closed$ \textbf{then} 

18 \textbf{\ \ \ \ \ \ \ \ \ if} $b'\notin open$ \textbf{then} 

19 $~\ \ \ \ \ \ \ \ \ \ c(b'):=\infty$; 

20 $~\ \ \ \ \ \ \ \ \ \ parent(b'):=NULL$; 

21 \ \ \ \ \ \ \ \ \ \textbf{end if}

22 $\ \ \ \ \ \ c_{old}:=c(b')$; 

23 \textbf{\ \ \ \ \ \ if} $c(parent(b))+\ln l(parent(b),b')<g(b')$
\textbf{then }

24 $~\ \ \ \ \ \ \ parent(b'):=parent(b)$; 

25 $\ ~\ \ ~\ \ \ c(b'):=c(parent(b))+\ln l(parent(b),b');$

26 \ \ \textbf{\ \ \ \ end if}

27 \textbf{\ \ \ \ \ \ if} $c(b')<c_{old}$ \textbf{then} 

28 \textbf{\ \ \ \ \ \ \ \ if} $b'\in open$ \textbf{then} 

29 $~\ \ \ \ \ \ \ open.Remove(b')$; 

30 \ \ \textbf{\ \ \ \ \ \ end if}

31 \emph{$~\ \ \ \ \ open.Insert(b',c(b')+h(b'))$}; 

32 \textbf{\ \ \ \ \ \ end if}

33 \ \ \ \ \ \textbf{end if}

34 \ \ \ \textbf{end for}

35 \textbf{\ \ end while}
\end{algorithm}

In the algorithm, $b_{BS}$ and $b_{user}$ are the start and goal
vertex of the search, respectively; $c(b)$ is the length of the shortest
path from the start vertex to $b$ found so far; $l(b',b)$ is the
straight-line distance between vertices $b$ and $b\prime$, and $LOS(b,b\prime)$
is true if and only if they have line of sight; $open.Insert(b,x)$
inserts vertex $b$ with key $x$ into the priority queue $open$;
$open.Remove(b)$ removes vertex $b$ from $open$; $open.Pop()$
removes a vertex with the smallest key from $open$ and returns it;
$parent(b)$ is used to extract the path at the end of the search;
$h(b)$ approximates the goal distance of the vertex $b$; $nghbr_{LOS}(b)$
is the set of neighbors of vertex $b$ that have line of sight to
$b$. 

Therefore, with the knowledge of the positions of the BS, users and
UAV-mounted STAR, joint 3D beam training can be carried out along
with channel estimation for each user.

\subsection{Complexity Analysis}

In this subsection, we compare the search complexity of our proposed
beam training schemes to those of others \cite{ning_terahertz_2021}.
The results are shown in Table \ref{tab:Comparison-of-beam}. Compared
to other schemes, the search time of our proposed scheme is negligible
and irrelevant to the total number of the STAR elements, $N^{2}$.
By contrast, the complexity of the benchmarks increases with the number
of the STAR elements.

\begin{table*}[tbh]
\caption{\label{tab:Comparison-of-beam}Comparison of beam training schemes}

\centering{}%
\begin{tabular}{|>{\centering}p{5cm}|>{\centering}p{3cm}|>{\centering}p{3cm}|>{\centering}p{5cm}|}
\hline 
\raggedright{}Beam training schemes &
Applicable to RIS-aided system &
Applicable to THz &
Search time for RIS-aided system\tabularnewline
\hline 
\hline 
\raggedright{}Exhaustive search &
Yes &
Yes &
$N^{2}+N^{4}$\tabularnewline
\hline 
\raggedright{}One-side search \cite{nitsche_ieee_2014} &
No &
No &
$-$\tabularnewline
\hline 
\raggedright{}Adaptive binary-tree search \cite{alkhateeb_channel_2014} &
No &
No &
$-$\tabularnewline
\hline 
\raggedright{}Two-stage training scheme \cite{lin_subarray-based_2017} &
No &
Yes &
$-$\tabularnewline
\hline 
\raggedright{}Tree dictionary (TD) and PS deactivation (PSD) codebook
based search \cite{ning_terahertz_2021} &
Yes &
Yes &
$18N+12\log_{3}N-3$ (TD) or $6N+4\log_{3}N-1$ (PSD)\tabularnewline
\hline 
\raggedright{}Proposed GI-based 3D beam training scheme  &
Yes &
Yes &
Negligible and not related to $N^{2}$\tabularnewline
\hline 
\end{tabular}
\end{table*}

\section{Simulation Results and Discussions}

In this section, we evaluate the performance of our Ghost Imaging-based
beam training and channel estimation scheme proposed for the UAV-mounted
STAR-aided THz multi-user MIMO systems. At the BS, each RF chain is
connected to a single sub-arrray and each subarray corresponds to
a user. As there is a low-attenuation THz transmission window at 350
GHz \cite{moldovan_and_2014}, we consider the THz frequency band
at 350 GHz in the simulations. The number of paths between the user
and the BS is set to 3 due to the sparsity of the THz channels \cite{lin_adaptive_2015}.
The BS and user positions are sensed by GI and the channel conditions
of (\ref{eq:angle}) are satisfied. The BS initially does not know
the CSI and the beam training and channel estimation is conducted
based on our proposed scheme. 

\begin{table}[tbh]
\caption{\label{tab:Simulation-Settings}Simulation Settings of the 5G Ghost
Imaging}

\centering{}%
\begin{tabular}{>{\raggedright}m{25mm}|>{\raggedleft}m{35mm}}
\hline 
Parameters &
Values\tabularnewline
\hline 
Duplex &
TDD\tabularnewline
\hline 
Central Frequency &
2.6 GHz\tabularnewline
\hline 
Sub-carriers Spacing &
15 KHz\tabularnewline
\hline 
CP Type &
Normal\tabularnewline
\hline 
\end{tabular}
\end{table}

\begin{table}[tbh]
\caption{\label{tab:Simulation-Parameters}THz 3D Beamforming Simulation Parameters}

\centering{}%
\begin{tabular}{>{\raggedright}m{25mm}|>{\centering}m{15mm}|>{\raggedleft}m{35mm}}
\hline 
Parameters &
Symbols &
Values\tabularnewline
\hline 
Center frequency of the subband &
$f_{h}$ &
$350$ GHz\tabularnewline
\hline 
Noise power &
$N_{0}$ &
\textminus 75 dBm\tabularnewline
\hline 
Subband Bandwidth &
$\varPi_{f_{h}}$ &
1 GHz\tabularnewline
\hline 
Absorption Coefficient  &
$\mu\left(f_{h}\right)$ &
2.13e-6 \cite{yao_stochastic_2017}\tabularnewline
\hline 
Number of Users &
$K$ &
4\tabularnewline
\hline 
Position of the center of the BS  &
$\mathbf{b}_{\mathit{BS}}$ &
(7, 13, 7)\tabularnewline
\hline 
Positions of the users in the focused cluster &
 &
(11, 8, 1), (13, 9, 0.94), (12, 8, 1.5), (14, 7, 0.8)\tabularnewline
\hline 
Position of the centroid of the focused user cluster &
$\mathbf{b}_{\mathit{user}}$ &
(12, 8, 1)\tabularnewline
\hline 
\end{tabular}
\end{table}

\begin{figure}[htbp]
\begin{centering}
\textsf{}%
\subfloat[]%
{\begin{centering}
\textsf{\includegraphics[scale=0.35]{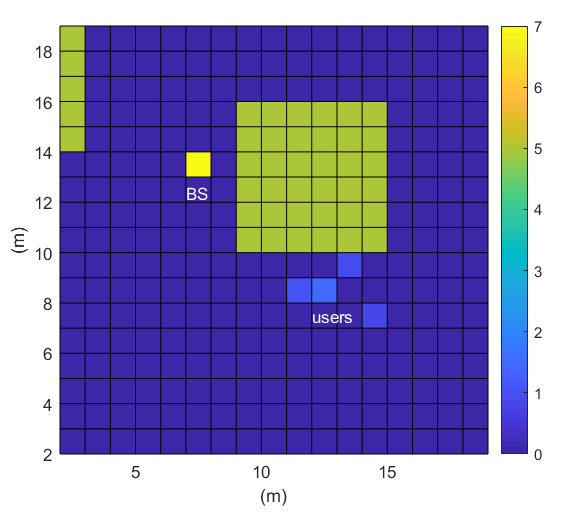}}
\par\end{centering}
}\textsf{ }%
\subfloat[]%
{\begin{centering}
\includegraphics[scale=0.35]{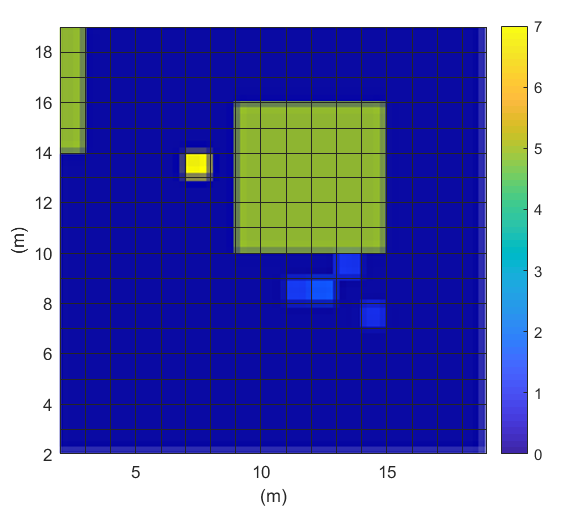}
\par\end{centering}
\textsf{}}
\par\end{centering}
\begin{centering}
\subfloat[]%
{\centering{}\includegraphics[scale=0.25]{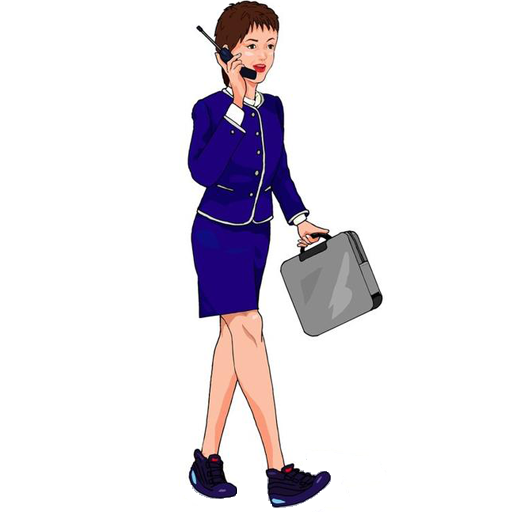}} %
\subfloat[]%
{\centering{}\includegraphics[scale=0.6]{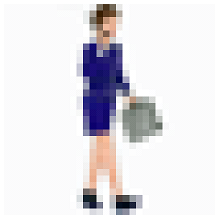}}
\par\end{centering}
\centering{}\caption{\label{fig:The-simulated-original}The originally investigated 3D
communication environment and reconstructed 5G Ghost Imaging results:
a) originally investigated communication environment; b) reconstructed
5G GI of the communication environment; c) original image of a user,
which can be regarded as the enlarged version of a user in (a); d)
5G GI reconstructed image of the user, which can be regarded as the
enlarged version of a user in (b). The color bars indicate the heights
of the objects in the image.}
\end{figure}

The simulation settings of the 5G GI are given in Table \ref{tab:Simulation-Settings}.
The originally investigated communication environment including the
users and the corresponding 5G GI reconstructed results are shown
in Fig. \ref{fig:The-simulated-original} (a)-(d), respectively. A
$20\times20\times20$ nodes grid representing a known cluttered environment
is created as shown in Fig. \ref{fig:The-simulated-original} (a).
The grid is populated with obstacles representing buildings in an
urban area. Location and size are randomly assigned to all obstacles.
The data representation of the environment is a simple $20\times20$
matrix with elements having discrete values from 0 to 20 m, which
is compatible with\emph{ Digital Elevation Models} (DEM), a 3D representation
of the terrain commonly adopted in topography. The color bars indicate
the heights of the objects in the image. The BS and users are also
specified in the image. The reconstructed 5G Ghost Imaging of the
communication environment is presented in Fig. \ref{fig:The-simulated-original}(b)
. The original image of a user, which can be regarded as the enlarged
version of the users in Fig. \ref{fig:The-simulated-original}(a),
and the corresponding reconstructed 5G GI image are shown in Fig.
\ref{fig:The-simulated-original}(c) and (d), respectively.

\begin{figure}[tbh]
\centering{}\includegraphics[scale=0.35]{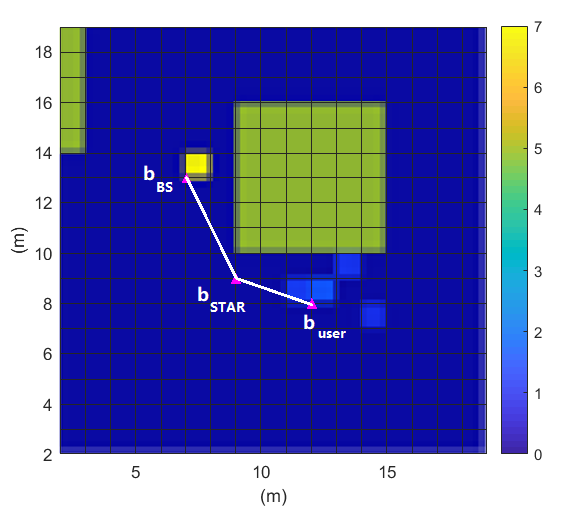}\caption{\label{fig:The-optimal-position-1}The optimal 3D position of the
UAV-mounted STAR based on Algorithm 1.}
\end{figure}

The 3D positions of the BS, users and centroid of the focused user
cluster are determined via GI and K-means, and shown in Table \ref{tab:Simulation-Parameters}.
For clarity, user clusters other than the focused one are not shown
in Fig. \ref{fig:The-simulated-original}. There are 4 users in the
focused cluster, at different positions. There are no LOS paths between
the BS and users. Based on these information and communication environment
including obstacles sensed by 5G GI, the optimal position of the UAV
STAR obtained by our proposed Algorithm 1 is (9, 9, 3), as shown in
Fig. \ref{fig:The-optimal-position-1}.

\begin{figure}[tp]
\begin{centering}
\textsf{}%
\subfloat[]%
{\textsf{\includegraphics[scale=0.25]{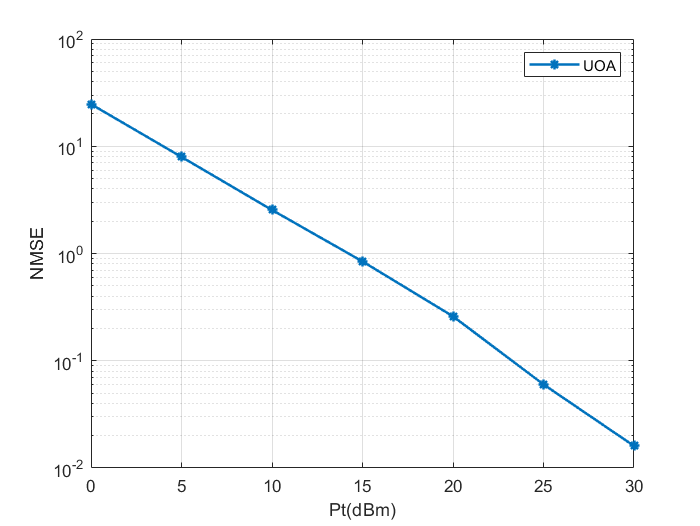}}} %
\subfloat[]%
{\includegraphics[scale=0.25]{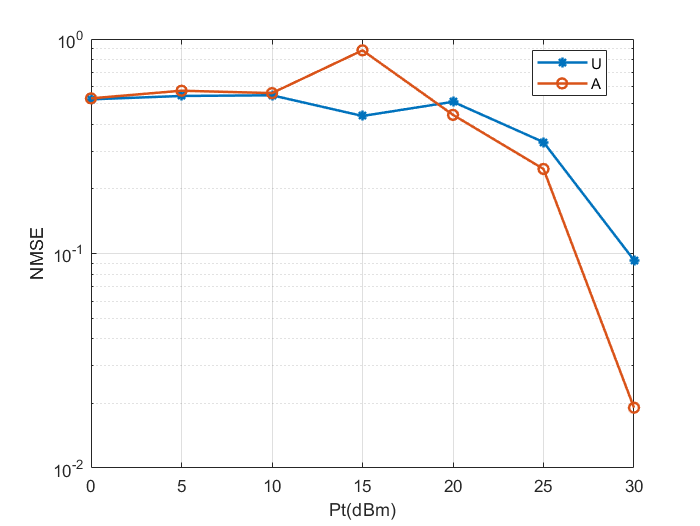}}
\par\end{centering}
\begin{centering}
\subfloat[]%
{\includegraphics[scale=0.4]{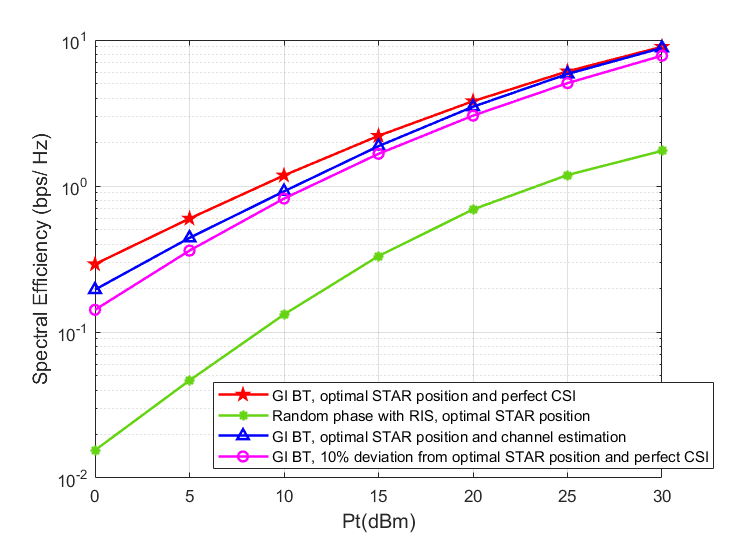}}
\par\end{centering}
\caption{\label{fig:Achievable-rate-(bps/Hz)}Joint 3D beamforming with 4 BS
UPA subarrays, 4 UPA subSTARs and 4 users: $4\times4$ UPA for each
BS subarray , $4\times4$ UPA for each subSTAR $4\times4$, and $4\times4$
UPA for each user. (a) Channel estimation of the cascaded channels;
(b) Channel estimation of the separated channels; (c) Achievable rate
(bps/Hz) versus transmit power (dBm).}
\end{figure}

The simulation parameters of the STAR-aided THz 3D beamforming for
multi-user massive MIMO systems are also given in Table \ref{tab:Simulation-Parameters}.
It is assumed that each BS antenna subarray is a $4\times4$ UPA and
each user is equipped with one RF chain associated with a $4\times4$
UPA array. Each subSTAR is also a $4\times4$ UPA. For the STAR-aided
channel, the distance between the BS and the STAR is 6 m while the
distances between the STAR and the users are 3 m, 4.5 m, 3.5 m and
5.8 m, respectively. The simulation results are shown in Fig. \ref{fig:Achievable-rate-(bps/Hz)}
(a)-(c). With the proposed semi-passive structure of STAR, the separated
channels are estimated efficiently as well as the cascaded channel.
Observe that the scheme with GI, the optimal STAR position and perfect
CSI achieves the highest spectral efficiency. The proposed scheme
combining GI and tensor based channel estimation together with the
optimal STAR position can effectively increase the spectral efficiency
with the increase of transmitting power. To demonstrate the advantage
of the proposed optimal STAR position find algorithm, we have chosen
a STAR position $\mathbf{b}_{dev}$ (8, 8, 2.1) which is 10\% deviated
from the optimal one but satisfies the rest of the constraints such
as $\mathbf{b}_{dev}\in\mathcal{F}$ and $||\mathbf{b}_{\mathit{BS}}-\mathbf{b}_{dev}||+||\mathbf{b}_{dev}-\mathbf{b}_{user}||>||\mathbf{b}_{BS}-\mathbf{b}_{user}||$.
The distance between the BS and the STAR is now 7.1 m while the distances
between the STAR and the users are 3.2 m, 5.2 m, 4.0 m and 6.2 m,
respectively. It can be seen that the spectral efficiency of the UAV-mounted
STAR at the deviated position even with perfect CSI is much lower
than that at the optimal position. The performance of STAR-aided channel
associated with random phase has the lowest spectral efficiency. 

\begin{figure}[tp]
\begin{centering}
\textsf{}%
\subfloat[]%
{\textsf{\includegraphics[scale=0.25]{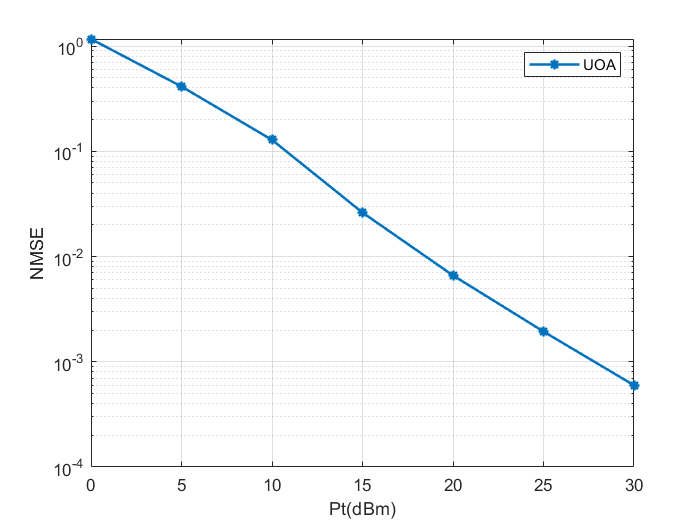}}} %
\subfloat[]%
{\includegraphics[scale=0.25]{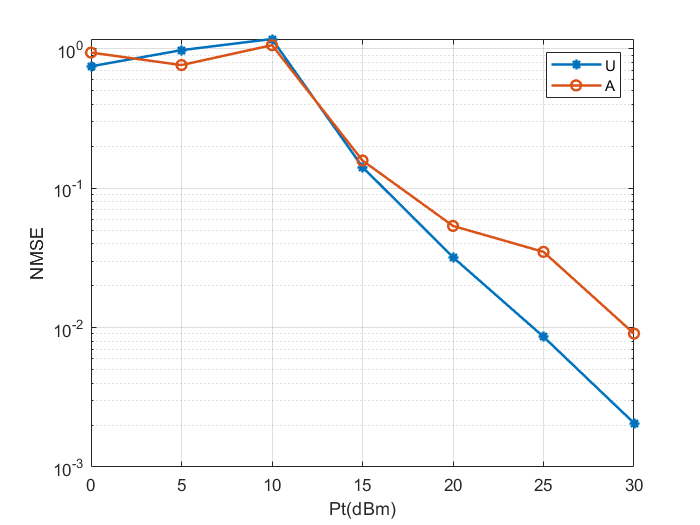}}
\par\end{centering}
\begin{centering}
\subfloat[]%
{\includegraphics[scale=0.4]{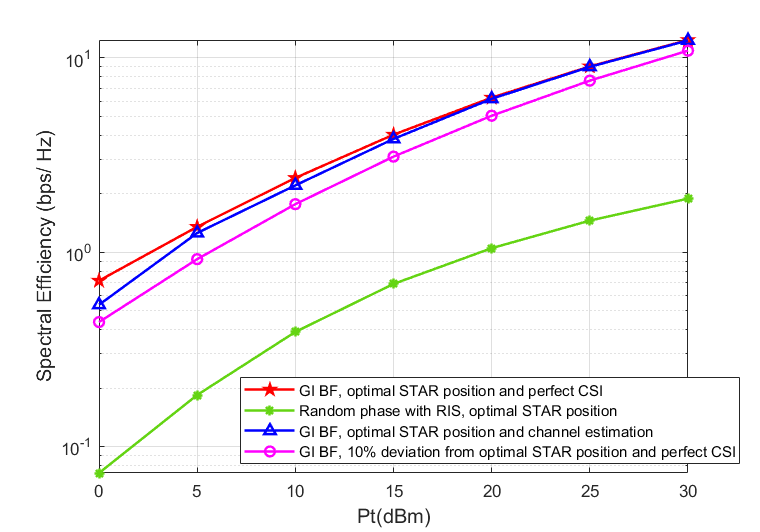}}
\par\end{centering}
\caption{\label{fig:Achievable-rate-(bps/Hz)-2}Joint 3D beamforming with 4
BS UPA subarrays, 4 UPA subSTARs and 4 users: $6\times6$ UPA for
each BS subarray , $6\times6$ UPA for each subSTAR, and $6\times6$
UPA for each user. (a) Channel estimation of the cascaded channels;
(b) Channel estimation of the separated channels; (c) Achievable rate
(bps/Hz) versus transmit power (dBm).}
\end{figure}

Then we increase the BS antenna subarray, STAR subarray and user array
all to $6\times6$ UPA. The simulation results are shown in Fig. \ref{fig:Achievable-rate-(bps/Hz)-2}.
It can be seen that with the increase of the STAR elements, the channel
estimation accuracy is further improved compared to the previous case.
Moreover, the system performance of the proposed scheme is improved
more significantly compared to that at the 10\% deviation position
even with perfect CSI.

\begin{figure}[tp]
\begin{centering}
\textsf{}%
\subfloat[]%
{\textsf{\includegraphics[scale=0.25]{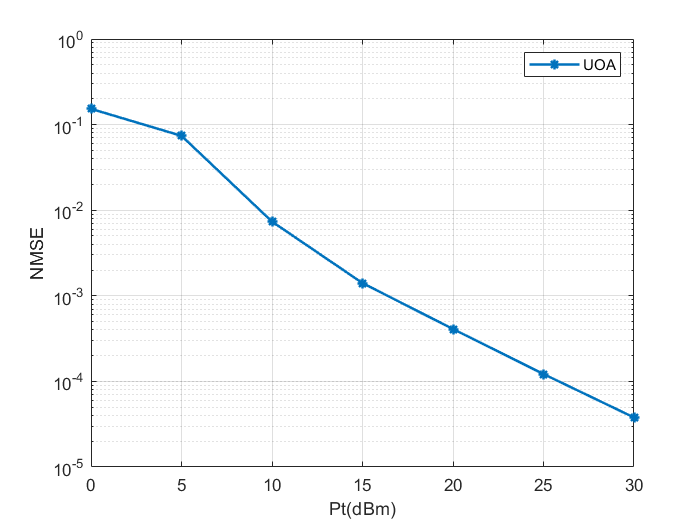}}} %
\subfloat[]%
{\includegraphics[scale=0.25]{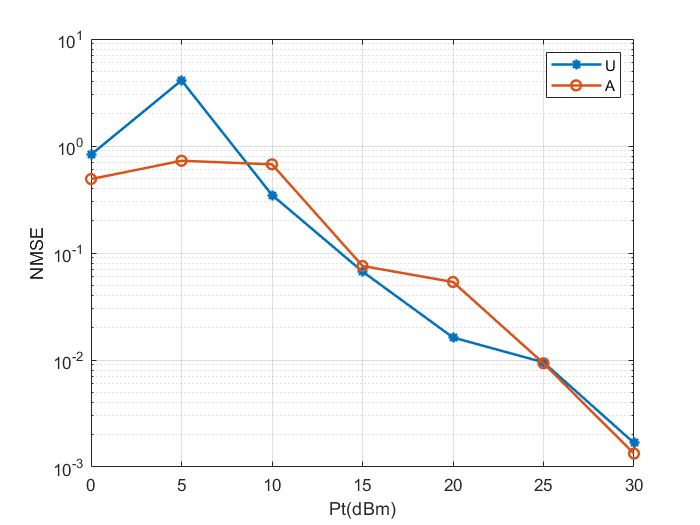}}
\par\end{centering}
\begin{centering}
\subfloat[]%
{\includegraphics[scale=0.4]{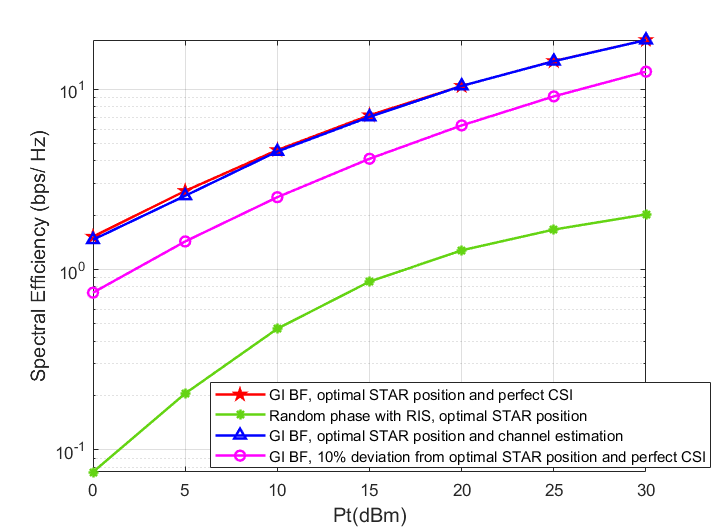}}
\par\end{centering}
\caption{\label{fig:Achievable-rate-(bps/Hz)-1}Joint 3D beamforming with 4
BS UPA subarrays, 4 UPA subSTARs and 4 users. $8\times8$ UPA for
each BS subarray , $8\times8$ UPA for each subSTAR, and $8\times8$
UPA for each user. (a) Channel estimation of the cascaded channel;
(b) Channel estimation of the separated channel; (c) Achievable rate
(bps/Hz) versus transmit power (dBm).}
\end{figure}

We continue to increase the BS antenna subarray, STAR subarray and
user array all to $8\times8$ UPA. The simulation results are shown
in Fig. \ref{fig:Achievable-rate-(bps/Hz)-1}. With the increase of
the STAR elements, the channel estimation accuracy is further improved
compared to the previous case. The system performance of the proposed
scheme is also improved, which is almost comparable to the perfect
CSI baseline. To sum up, the proposed scheme can improve the spectral
efficiency on average by 14.26\%, 18.35\% and 60.60\% for the three
configurations, respectively, compared to that at the 10\% deviation
position even with perfect CSI.

\section{Conclusions}

A novel Ghost Imaging-based joint 3D beam training scheme was proposed
in this paper for UAV-mounted STAR aided THz multi-user massive MIMO
systems. To maximize the average achievable sum-rate of the users,
the proposed BS sub-array and sub-STAR spatial multiplexing architecture,
optimal active and passive beamforming, digital precoding, and optimal
position of the UAV-mounted STAR were jointly investigated. The surrounded
5G downlink signals are utilized for GI. Our simulation results show
that the spectral efficiency of our proposed system is much higher
than that at the deviated position or with the STAR at random phases.
Our proposed GI based joint 3D beamforming architecture with the optimal
position finding algorithm has practical potential for emerging UAV-mounted
STAR-aided THz applications such as integrated networks of terrestrial
links, UAVs, and satellite communication systems. As our future work,
other forms of ISAC will be explored for UAV-mounted STAR-aided THz
multi-user massive MIMO systems.

\section{Appendix A\protect \\
Ambiguity Elimination for Separation Channel Estimation}

When combined with tensor-based channel estimation, the received signal
for user \emph{k} can be expressed by

\begin{equation}
\tilde{\mathbf{y}}_{k}[t]=\mathbf{v}_{k}^{H}\mathbf{H}_{k}\mathbf{\mathbf{O_{\mathit{k}}^{\mathit{TR}}}}\mathbf{G}_{k}\mathbf{\mathbf{W_{\mathit{k}}}}\mathbf{F_{\mathit{k}}\mathbf{x_{\mathit{k}}}}[t]+\mathbf{v}_{k}^{H}\mathbf{n}_{k}[t],\quad1\leq t\leq T.\label{eq:tensor}
\end{equation}

\noindent where $\mathbf{x}_{k}[t]\in\mathbb{C}^{m_{t}n_{t}\times1}$
is the vector of the transmitted pilot signals for the user \emph{k}
at time $t$. The channel training time $T_{s}$ is divided into $B$
blocks, where each block has $T$ time slots so that $T_{s}=BT$.
The signal part of the equation (\ref{eq:tensor}) can be expressed
by 

\begin{equation}
\mathbf{\overline{Y}_{\mathit{s}}^{\mathit{k}}}[i]=\mathbf{U^{\mathit{k}}}\mathbf{O}_{i}^{k}\mathbf{Z}_{k}^{\mathrm{T}}=\mathbf{U^{\mathit{k}}}\mathbf{O}_{i}^{k}\mathbf{A^{\mathit{k}}}\mathbf{X_{\mathit{k}}^{\mathrm{T}}},~i=1,...,B
\end{equation}
where $\mathbf{O}_{i}^{k}\doteq\mathrm{diag}(\mathbf{q}[i])$ denotes
a diagonal matrix holding the $i$-th row of the STAR phase shift
matrix $\mathbf{O^{\mathit{k}}}$ on its main diagonal, $\mathbf{\mathbf{Z^{\mathit{k}}}}=\mathbf{X^{\mathit{k}}A_{\mathit{k}}^{\mathrm{T}}}$,
$\mathbf{A^{\mathit{k}}}=\mathbf{G^{\mathit{k}}W^{\mathit{k}}F^{\mathit{k}}}$,
$\mathbf{U_{\mathit{k}}}=\mathbf{V}_{\mathit{k}}^{\mathrm{H}}\mathbf{H_{\mathit{k}}}$
and $\mathbf{X^{\mathit{k}}}\in\mathbb{C}^{T\times L_{B}}$ are the
$k$-th user pilot matrix, respectively. The matrix $\mathbf{\overline{Y}_{\mathit{s}}^{\mathit{k}}}[i]$
can be viewed as the $i$-th frontal matrix slice of a three-way tensor
$\overline{\mathcal{Y}}^{k}\in\mathbb{C}^{L_{B}\times T\times B}$
following a PARAFAC decomposition.

As long as the sufficient conditions \cite{de_araujo_channel_2021}
are satisfied , the uniqueness of the nested PARAFAC decomposition
means that two quintuplets of matrix factors are linked by

\begin{equation}
\overline{\mathbf{U}}^{k}=\mathbf{U}^{k}\boldsymbol{\Pi}\boldsymbol{\Lambda}^{U},\label{eq:R}
\end{equation}
\begin{equation}
\overline{\mathbf{O}}^{k}=\mathbf{O^{\mathit{k}}}\boldsymbol{\Pi}\boldsymbol{\Lambda}^{O},\label{eq:O}
\end{equation}

\begin{equation}
\overline{\mathbf{A}}^{k}=\mathbf{A}^{k}\boldsymbol{\Pi}\boldsymbol{\Lambda}^{A},\label{eq:A}
\end{equation}
where $\boldsymbol{\Lambda}^{U}$, $\boldsymbol{\Lambda}^{O}$ and
$\boldsymbol{\Lambda}^{A}$ are diagonal matrices, and $\boldsymbol{\Pi}$
is a permutation matrix. Taking into account the knowledge of the
phase shift matrices $\mathbf{O^{\mathit{k}}}$ of the subSTAR $k$,
the ambiguity relation (\ref{eq:O}) gives $\boldsymbol{\Pi}=\boldsymbol{\Lambda}^{O}=\mathbf{I}_{O}$,
then (\ref{eq:R}) and (\ref{eq:A}) become 

\begin{equation}
\overline{\mathbf{U}}^{k}=\mathbf{U}^{k}\boldsymbol{\Lambda}^{U},
\end{equation}

\begin{equation}
\overline{\mathbf{A}}^{k}=\left(\boldsymbol{\Lambda}^{U}\right)^{-1}\mathbf{A}^{k}.
\end{equation}

The only remaining ambiguities exist in the diagonal matrices $\boldsymbol{\Lambda}^{U}$
or $\boldsymbol{\Lambda}^{A}$. If we know any row of $\mathbf{U}^{k}$
or $\mathbf{A^{\mathit{k}}}$, e.g., the $z$-th row, these scaling
ambiguities can then be eliminated by the following operations: 

\begin{equation}
\widehat{\boldsymbol{\Lambda}}^{U}=\mathcal{L}_{z}\left(\mathbf{U}^{k}\right)\mathcal{L}_{z}^{-1}\left(\widehat{\mathbf{U}}^{k}\right),\label{eq:e1}
\end{equation}

\begin{equation}
\widehat{\boldsymbol{\Lambda}}^{A}=\mathcal{L}_{z}(\mathbf{A}^{k})\mathcal{L}_{z}^{-1}\left(\widehat{\mathbf{A}}^{k}\right),\label{eq:e2}
\end{equation}
where $\widehat{\mathbf{U}}^{k}$ and $\widehat{\mathbf{A}}^{k}$
denote the estimated values of $\mathbf{U}^{k}$ and $\mathbf{\mathbf{\mathbf{A^{\mathit{k}}}}}$
at convergence of the BALS algorithms, $\mathcal{L}_{z}(\mathbf{A})$
denotes the the $z$-th row of matrix $\mathbf{A}$.

Compare to the cascaded channel estimation cases, we have the remark
as follows:

\noindent \textbf{Remark 1: The ambiguity in the separated channel
$\mathbf{U}^{k}$ and $\mathbf{\mathbf{\mathbf{A^{\mathit{k}}}}}$
may affect the optimal passive beamforming design.}

According to (\ref{eq:R}) to (\ref{eq:A}), if the phase shift matrices
$\mathbf{O^{\mathit{k}}}$ of the subSTAR are unknown, then any phase
shift matrices satisfied

\begin{equation}
\overline{\mathbf{O}}^{k}=\left(\boldsymbol{\Lambda}^{U}\right)^{-1}\mathbf{O^{\mathit{k}}}\left(\boldsymbol{\Lambda}^{A}\right)^{-1},
\end{equation}
can also be the pursued phase shift matrices. Therefore, it is necessary
to eliminate the ambiguities in the channel $\mathbf{U}^{k}$ and
$\mathbf{\mathbf{\mathbf{A^{\mathit{k}}}}}$. 

Here we propose a semi-passive STAR architecture with each subSTAR
being composed of one special STAR element equipped with a receiving
RF chain, which can operate in both sensing and reflecting / transmitting
modes as illustrated in Fig. \ref{fig:System-model-of}. The special
STAR element is able to directly sense the one-hop channels by observing
the training signal at the STAR end. Resorting to the special STAR
elements, we will be able to obtain one of the rows of $\mathbf{U}^{k}$
or $\mathbf{A^{\mathit{k}}}$, e.g., the $z$-th row. According to
(\ref{eq:e1}) and (\ref{eq:e2}), the ambiguities in the channel
$\mathbf{U}^{k}$ and $\mathbf{\mathbf{\mathbf{A^{\mathit{k}}}}}$
can then be eliminated. 

% \bibliographystyle{IEEEtran}
% \bibliography{IEEEabrv,20D__Australia_thz_meeting_codetens_paper________________________________}


\end{document}